\documentclass[twocolumn,superscriptaddress,aps,pre,amssymb,amsfonts,amsmath]{revtex4-1}
\usepackage{bm}
\usepackage{graphicx}
\usepackage{xcolor}
\newcommand{\rev}[1]{\textcolor{black}{{#1}}}
\definecolor{darkblue}{rgb}{0,0,0.6}
\usepackage[colorlinks,linkcolor=darkblue,citecolor=darkblue,urlcolor=darkblue]{hyperref}

\begin{document}

\title{Static self-induced heterogeneity in glass-forming liquids: Overlap as a microscope}

\author{Benjamin Guiselin}

\affiliation{Laboratoire Charles Coulomb (L2C), Universit\'e de Montpellier, CNRS, 34095 Montpellier, France}

\author{Gilles Tarjus}

\affiliation{LPTMC, CNRS-UMR 7600, Sorbonne Universit\'e, 4 Pl. Jussieu, F-75005 Paris, France}

\author{Ludovic Berthier}

\affiliation{Laboratoire Charles Coulomb (L2C), Universit\'e de Montpellier, CNRS, 34095 Montpellier, France}

\affiliation{Yusuf Hamied Department of Chemistry, University of Cambridge, Lensfield Road, Cambridge CB2 1EW, United Kingdom}

\email{ludovic.berthier@umontpellier.fr}

\date{\today}

\begin{abstract}
We propose and numerically implement a local probe of the static self-induced heterogeneity characterizing glass-forming liquids. The method relies on the equilibrium statistics of the overlap between pairs of configurations measured in mesoscopic cavities with unconstrained boundaries. By systematically changing the location of the probed cavity, we directly detect \rev{spatial variations of the overlap fluctuations.} We provide a detailed analysis of the statistics of a local estimate of the configurational entropy and we infer an estimate of the surface tension between amorphous states, ingredients that are both at the basis of the random first-order transition theory of glass formation. \rev{Our results represent the first direct attempt to visualize and quantify the self-induced heterogeneity underpinning the thermodynamics of glass formation. They pave the way for the development of coarse-grained effective theories and for a direct assessment of the role of thermodynamics in the activated dynamics of deeply supercooled liquids.} 
\end{abstract}
\maketitle

\section{Introduction}
\label{sec:intro}

Glass formation is a universal phenomenon resulting from the rapid increase of the viscosity or the structural relaxation time $\tau_\alpha(T)$ of supercooled liquids 
when lowering the temperature~\cite{berthier2011theoretical}. The viscosity eventually becomes so large that the liquid no longer flows on experimental timescales and behaves as a nonequilibrium amorphous solid, {\it i.e.}, a glass: this operationally defines the glass transition temperature $T_\mathrm{g}$. This slowing down comes with an increasing heterogeneity of the dynamics which is now well characterized in experiments and computer simulations~\cite{berthier2011overview, berthier2011}. At low enough temperature, relaxation is not spatially homogeneous and the sample rather contains domains of high and low mobility with a typical lifetime of the order of $\tau_\alpha(T)$ and a size characterized by the dynamic correlation length $\xi_\mathrm{d}(\tau_\alpha(T))$. The latter grows upon decreasing the temperature $T$~\cite{berthier2005direct, dalle2007spatial}.

The Random First-Order Transition (RFOT) theory, first developed by Kirkpatrick, Thirumalai and Wolynes~\cite{kirkpatrick1989scaling}, describes glass formation in finite dimensions $d$ by building on the mean-field picture of glassiness governed by the properties of an underlying rugged free energy 
landscape~\cite{lubchenko2007theory,wolynes2012structural,parisi2020theory}. Relaxation slowdown is controlled by the approach to a thermodynamic glass transition (the RFOT) at a temperature $T_\mathrm{K}<T_\mathrm{g}$ which is characterized by a diverging lengthscale, $\xi_\mathrm{ps}(T)$. This static correlation length has been identified as the point-to-set correlation length and represents the average linear size $R$ over which the density profile of the liquid at a given point $\bm{x}$ is constrained by the set of particles at a distance $R$ from it~\cite{bouchaud2004adam, biroli2012random, montanari2006rigorous}. It can be interpreted as resulting from the competition between a bulk free energy gain of entropic nature that comes from the possibility to explore an exponentially large number of accessible amorphous states and a surface free energy cost associated with the coexistence of two different amorphous states~\cite{bouchaud2004adam}. The former is given by the configurational entropy per particle,  $\Sigma(T)$, which decreases with decreasing temperature $T$ until, at least in the mean-field scenario, it vanishes at $T_\mathrm{K}$. The latter is associated with a generalized surface tension $\Upsilon(T)$. Due to this free energy competition the liquid is assumed to appear, at low enough temperature but still above  $T_\mathrm{K}$, in a `mosaic state'. This can be schematically thought of as the juxtaposition of different amorphous states extending over a typical linear size~\cite{kirkpatrick1989scaling} 
\begin{equation}
\xi_\mathrm{ps}(T)\sim \left[\frac{\Upsilon(T)}{\rho T\Sigma(T)}\right]^{1/(d-\theta)},
\label{eqn:pts}
\end{equation}
with $\theta\leq d-1$~\cite{kirkpatrick1989scaling, franz2011analytical} and $\rho$ the number density.

The RFOT theory then describes the structural relaxation of the liquid via the thermally activated dynamics of the mosaic domains~\cite{kirkpatrick1989scaling, lubchenko2007theory,xia2000fragilities} with a typical free energy barrier scaling as $[\xi_\mathrm{ps}(T)]^\psi$, where $\psi\leq d-1$. This description naturally captures the dynamical slowdown with $\ln\tau_\alpha(T)\sim [1/\Sigma(T)]^{\psi/(d-\theta)}$. The mosaic picture also rationalizes the heterogeneous nature of the dynamics at low temperature. 

The existence of a complex free energy landscape with a multitude of metastable states, which is found at the mean-field level~\cite{castellani2005spin, parisi2020theory} and is postulated in finite dimensions by the RFOT theory, leads to nontrivial thermodynamic fluctuations of the overlap order parameter, $\widehat Q[\bm{r}^N;\bm{r}_0^N]$, which represents the degree of similarity between the liquid configuration, $\bm{r}^N$, and a reference configuration $\bm{r}_0^N$ of the same liquid. As a result, equilibrium phase transitions are expected when the global overlap is linearly coupled to a field $\epsilon$~\cite{franz1997phase} or when a fraction $c$ of particles are pinned~\cite{cammarota2012ideal, kob2013probing, cammarota2013random}, with associated critical points  in the universality class of the random-field Ising model (RFIM)~\cite{franz2013universality, biroli2014random}.

The predictions of the RFOT approach for the dynamics remain difficult to precisely assess beyond a general qualitative agreement with the phenomenology of glass-forming liquids and empirical correlations~\cite{lubchenko2007theory, bouchaud2004adam, tarjus2011overview, ozawa2019does}. By contrast, the predictions for the statics have received substantial support from numerical studies. Generally speaking, the fluctuations of the overlap order parameter are found to behave in small systems as in the mean-field theory~\cite{guiselin2021statistical, franz1997phase, parisi2014liquid,cammarota2010phase, berthier2013overlap, kob2013probing,berthier2014novel, cammarota2016first, berthier2015evidence, jack2016phase}. It has also been shown that a point-to-set length can indeed be measured by considering cavities with frozen boundaries~\cite{bouchaud2004adam, biroli2012random} and mildly grows upon decreasing the temperature~\cite{cavagna2007mosaic, biroli2008thermodynamic, nagamanasa2015direct, yaida2016point, berthier2016efficient, berthier2019zero}, and, through finite-size scaling, evidence has been provided for the existence of a transition in the presence of an applied field $\epsilon$ that terminates in a critical point in the RFIM universality class~\cite{berthier2015evidence, guiselin2020random, guiselin2021statistical}. In addition, signatures of a nonzero surface tension between amorphous states have been obtained~\cite{cammarota2009evidence, cammarota2009numerical, ganapathi2018measurements} and the configurational entropy has been measured by a variety of techniques~\cite{berthier2019configurational} which all report a modest decrease with decreasing temperature~\cite{sastry2001relationship, sengupta2012adam, berthier2014novel, berthier2017configurational, ozawa2018configurational, berthier2019zero}. Note, however, that the connection between the measured configurational entropies and the mean-field construct is far from trivial: see, {\it e.g.}, Refs.~[\onlinecite{berthier2019configurational}, \onlinecite{cammarota2011renormalization}]. 

\rev{This accumulation of results concerning the macroscopic thermodynamical behavior of glass-formers provides encouraging signs that the RFOT theory is a solid starting point. Despite this important progress, many open questions remain in connection with the RFOT theory. The very notion of a mosaic picture requires going beyond macroscopic evidence and deals with local scale fluctuations. Our work is an effort in this direction. In particular, the real-space characterization of the mosaic itself remains rather fuzzy, to say the least~\cite{cammarota2012patch}.} Next, the configurational entropy, the surface tension, and the point-to-set length, are all expected to be random variables~\cite{xia2001microscopic, lubchenko2004theory, dzero2009replica, biroli2012random} that fluctuate in space. These fluctuations, which are associated with some kind of static heterogeneity of glass-forming liquids, can be interpreted as resulting from a \rev{self-induced disorder~\cite{bouchaud1994self}. This terminology comes from the quantitative analogy, at the mean-field level, between liquids in infinite dimensions and fully-connected spin glass models which both exhibit a rough free-energy landscape at the origin of their glassy slowdown. For the latter, the quenched disordered interactions introduced in the Hamiltonian are directly responsible for the emergence of a rugged landscape. In supercooled liquids, particle interactions are not random but frustration nevertheless leads to similar complex free-energy landscapes which are then an emerging physical property: hence the term `self-induced'. This should not be confused with the more obvious static heterogeneity emerging from the aperiodic nature of liquid configurations.}

\rev{The self-induced disorder} is responsible for the RFIM universality class found for overlap fluctuations and, accordingly, they imply that $T_\mathrm{K}=0$ in $2d$ and that a $T_\mathrm{K}>0$ may exist in $3d$ only if these fluctuations are weak enough~\cite{biroli2014random}. The characteristics of the self-induced disorder are then important to assess whether a thermodynamic glass transition exists in $3d$ glass-forming liquids or, more ambitiously, to build an effective theory of the glass transition with parameters directly obtained from actual supercooled liquids~\cite{stevenson2008constructing, biroli2018random1, biroli2018random2}. Roughly speaking, the spatial fluctuations of the configurational entropy are the source of the emergent random field and those of the surface tension the source of an emergent random-bond disorder~\cite{stevenson2008constructing, biroli2014random, biroli2018random1, biroli2018random2}. \rev{It would be desirable to have direct access to these fluctuations.}

Finally, one would of course like to make a causal connection between the static properties associated with the overlap fluctuations, which all seem to be in qualitative agreement with the mean-field and RFOT theory approaches, and the salient dynamical phenomena observed in glass-forming liquids, super-Arrhenius activated relaxation, spatially heterogeneous dynamics, nonexponential behavior of the time-dependent correlation functions, etc. Our contention is that making empirical correlations between static and dynamic quantities more significant, and therefore more indicative of an actual causal relationship, requires to go beyond the investigation of global correlations and to study local ones (see also Ref.~[\onlinecite{PhysRevLett.127.088002}]). Local here refers to a mesoscopic scale over which thermodynamic-like quantities such as a local configurational entropy can be reasonably defined rather than a purely microscopic, particle-based, one.

\rev{The primary objective of this work is to provide the first steps in these directions and to shift the analysis of thermodynamic fluctuations toward a more local scale. To this end, we develop} a new probe to directly assess the self-induced static heterogeneity in glass-forming liquids via local free energy measurements. By free energy, we mean the setup of the Franz-Parisi potential~\cite{franz1997phase} which characterizes the cost of keeping liquid configurations at a given overlap with a reference configuration of the same liquid.  We also introduce a field $\epsilon$ to bias the overlap with a reference configuration $\bm{r}_0^N$ in a spherical cavity of radius $R$ while letting the outside of the cavity evolve without any thermodynamic constraint. By varying the radius of the cavity and by systematically changing the location of the cavity, we are in particular able to directly probe the spatial heterogeneity of the configurational entropy density and we can analyze its statistics and spatial organization. This gives a real-space description of the emergent local random field and provides the necessary ingredients for a better understanding of the mosaic picture proposed by the RFOT theory. \rev{Our results show that all these quantities can be defined and accessed numerically, which was not guaranteed by the success of macroscopic measurements. Building on our approach, we believe that connections to structural relaxation can be addressed in the future, thus paving the way to answer several important questions regarding the RFOT theory.}

In several previous attempts to characterize the structural heterogeneity of supercooled liquids, the system needs to first be quenched to its inherent state at zero temperature where some kind of structural property, related to mechanical moduli~\cite{mizuno2013measuring, shakerpoor2020stability}, harmonic excitations~\cite{widmer2008irreversible}, linear~\cite{doi:10.1063/1.5024776} or non-linear~\cite{barbot2018local} response to a localised perturbation, is analyzed. By construction, the connection with finite temperature physical properties is at best indirect. A second family of structural studies relies on the analysis of finite-temperature equilibrium states and mostly uses particle-based geometric information to reveal spatial fluctuations about the local ordering of the liquid~\cite{coslovich2011locally, malins2013identification1, malins2013identification, tong2018revealing}, an approach that is now assisted  by machine learning techniques~\cite{paret2020assessing, boattini2020autonomously}. It remains difficult to incorporate these findings into a generic thermodynamic approach accounting for the observed fluctuations and phase transitions described above.

The remaining of the article is organized as follows. In Sec.~\ref{sec:stat_mech}, we discuss several settings that can in principle give access to the spatial fluctuations of the configurational entropy density and of related quantities, reviewing in particular the role played by boundary conditions. The liquid model and the numerical methods are presented in Sec.~\ref{sec:methods}. We describe our results and show illustrative maps of the self-induced disorder in Sec.~\ref{sec:results}. We finally provide conclusions and perspectives in Sec.~\ref{sec:ccl}.

\section{Strategies to measure local overlap fluctuations}
\label{sec:stat_mech}

In order to access the self-induced disorder \rev{characterizing} glass-forming liquids, one needs to consider the overlap fluctuations of mesoscopic subsystems and to characterize their variations from one position to another. We discuss three different strategies, corresponding to different boundary conditions applied on the subsystem, which can be implemented toward this goal. These different geometries are sketched in Fig.~\ref{fig:settings}.

\subsection{Point-to-set construction with frozen boundaries}
\label{sec:stat_mech_pts}

\begin{figure}[t]
\includegraphics[width=0.49\columnwidth]{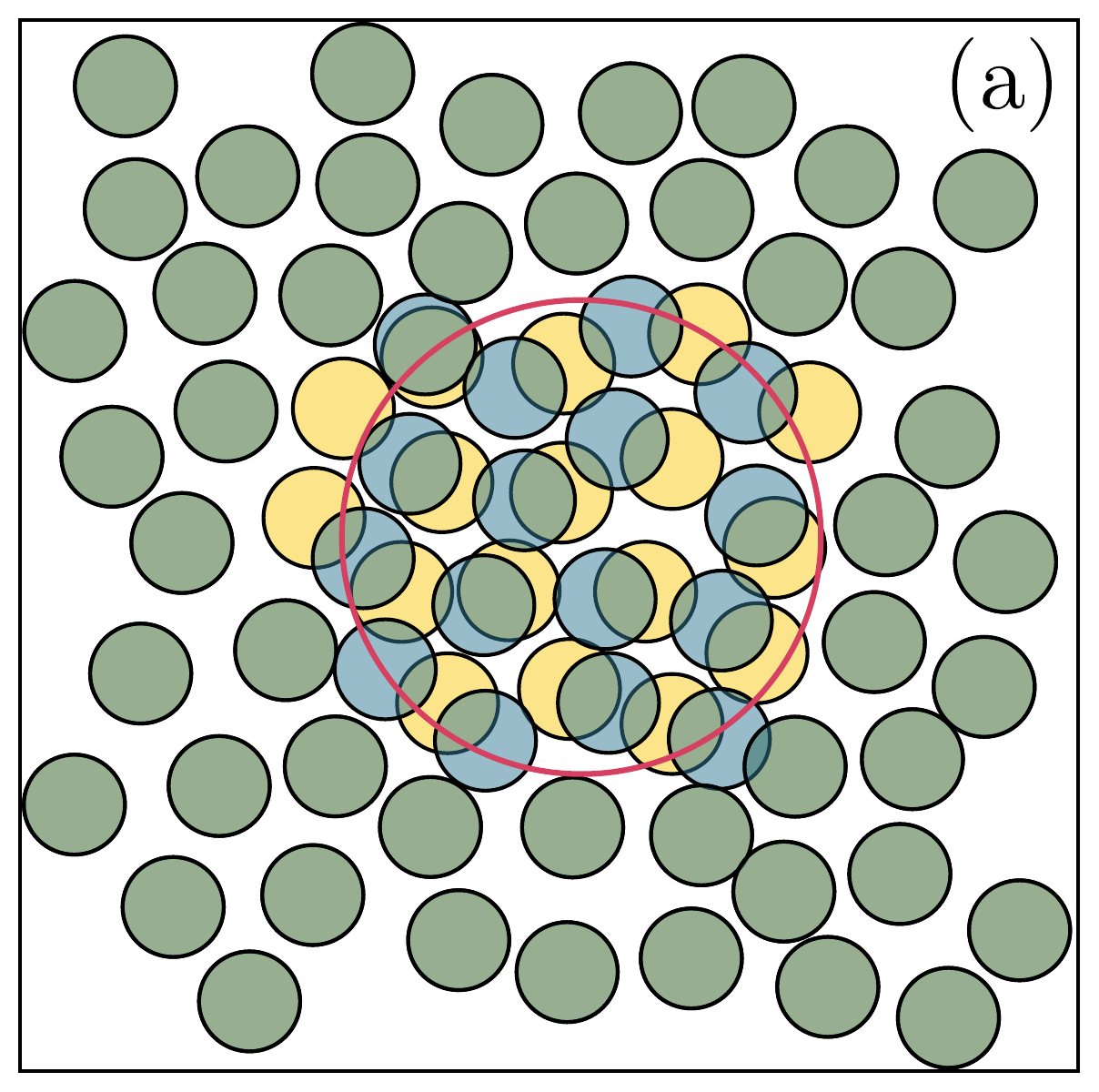}
\includegraphics[width=0.49\columnwidth]{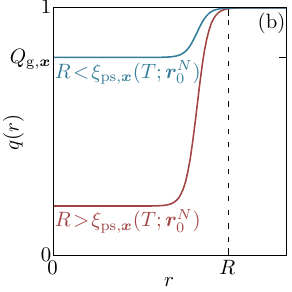}
\includegraphics[width=0.49\columnwidth]{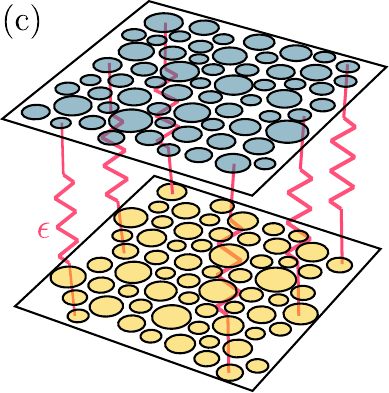}
\includegraphics[width=0.49\columnwidth]{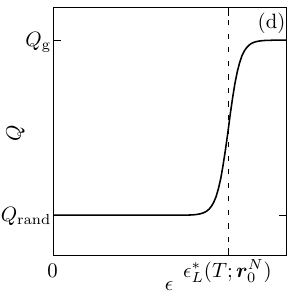}
\includegraphics[width=0.49\columnwidth]{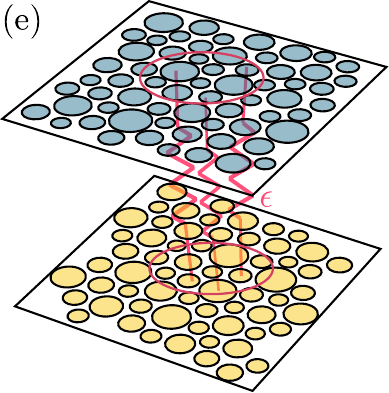}
\includegraphics[width=0.49\columnwidth]{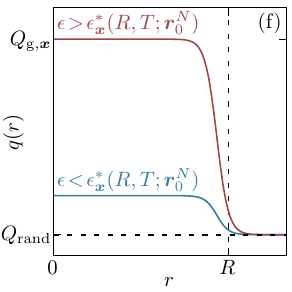}
\caption{Sketch of three different settings to investigate the self-induced disorder in supercooled liquids at the mesoscopic scale. (a, b) Point-to-set construction: a cavity of radius $R$ at position $\bm{x}$ explores the configuration space in the presence of a pinning boundary drawn from an equilibrium configuration $\bm{r}_0^N$. The overlap at the center of the cavity significantly decreases when $R$ reaches the local point-to-set length $\xi_{\mathrm{ps},\bm{x}}(T;\bm{r}_0^N)$. (c, d) Mesoscopic system with periodic boundary conditions: the global overlap $\widehat Q$ with a reference configuration is linearly biased by an applied field $\epsilon$ and its thermal average increases significantly when $\epsilon$ reaches the crossover field $\epsilon_{L}^*(T;\bm{r}_0^N)$. (e, f) This work: The overlap $\widehat Q_{\bm{x}}^{(R)}$ in a cavity of radius $R$ is biased with a field $\epsilon$ acting only inside the cavity. The overlap at the center of the cavity significantly increases when $\epsilon$ reaches the crossover field $\epsilon^*_{\bm{x}}(R,T;\bm{r}_0^N)$. In (b, d, f), $Q_\mathrm{rand}\ll 1$ stands for the overlap between  uncorrelated configurations, while $Q_\mathrm{g}\approx 1$ stands for the overlap between nearby configurations (which may also fluctuate).}
\label{fig:settings}
\end{figure}

The point-to-set construction relies on the study of the statistical mechanics at a temperature $T$ of a cavity of radius $R$ and position $\bm{x}$ in a frozen environment drawn from a reference equilibrium configuration $\bm{r}_0^N$~\cite{bouchaud2004adam, biroli2012random}, see Fig.~\ref{fig:settings}(a). This amounts to studying the thermodynamics of a mesoscopic cavity in the presence of a pinning boundary characterized by a high overlap with the reference configuration. This boundary condition induces an inhomogeneous overlap profile, $q(r)$, converging to 1 (or a high value if one does not constrain the exterior of the cavity to be exactly frozen in the reference configuration) when $r \geq R$, see Fig.~\ref{fig:settings}(b). 

Within the mean-field and RFOT theory settings, the thermodynamic state of the cavity, as characterized by its overlap with its counterpart in the reference configuration, results from the competition between a configurational entropy gain, $-T\Sigma_{\bm{x}}^{(R)}(T;\bm{r}_0^N) \rho V_d R^d$, which drives the cavity to explore different amorphous states and a surface free energy cost, $\Upsilon_{\bm{x}}^{(R)}(T;\bm{r}_0^N) S_d R^\theta$, when there is a mismatch in density profiles at the boundary of the cavity. Here,  $V_d$ and $S_d=dV_d$ are the volume and the area of the unit sphere in $d$ dimensions, and $\theta\leq d-1$. The value $\theta=d/2$ was proposed from a wetting argument~\cite{kirkpatrick1989scaling}, while $\theta=d-1$ was obtained in simulations~\cite{cammarota2009evidence, cammarota2009numerical, ozawa2019does}, by considering a model spin glass with large but finite-range interactions~\cite{franz2011analytical} and through instanton calculations~\cite{franz2005first, dzero2005activated}. The total free energy per particle when the cavity is in a different amorphous state than the reference configuration then reads  
\begin{equation}
\Delta F_{\bm{x}}(R,T;\bm{r}_0^N)=-T\Sigma_{\bm{x}}^{(R)}(T;\bm{r}_0^N)+\frac{d}{\rho R^{d-\theta}}\Upsilon_{\bm{x}}^{(R)}(T;\bm{r}_0^N).
\end{equation}
The coarse-grained configurational  entropy density $\Sigma^{(R)}_{\bm{x}}(T;\bm{r}_0^N)$ stands for the average of $\Sigma_{\bm{x}}(T;\bm{r}_0^N)$ over the cavity of radius $R$ centered at position $\bm{x}$. It crosses over from $\Sigma_{\bm{x}}(T;\bm{r}_0^N)$ for $R\ll \xi_\Sigma(T)$ to the average configurational entropy density $\Sigma(T)= \langle {\Sigma_{\bm{x}}(T;\bm{r}_0^N)}\rangle$ for $R\gg \xi_\Sigma(T)$, where $\xi_\Sigma(T)$ is the typical correlation length of the configurational entropy and the brackets denote an average over the positions of the cavity 
and over the reference configurations. The local surface tension $\Upsilon_{\bm{x}}^{(R)}(T;\bm{r}_0^N)$ is expected to depend on the density profile in the reference configuration at the boundary with the cavity, resulting in a dependence on both $\bm{x}$ and $R$. All these quantities depend on the reference configuration $\bm{r}_0^N$.

Within this point-to-set construction with frozen boundaries, the radius $R$ of the cavity plays the role of a control parameter and the overlap at the center of the cavity crosses over from a high value for $R<\xi_{\mathrm{ps},\bm{x}}(T;\bm{r}_0^N)$ ($\Delta F_{\bm{x}}>0$) when the state in the cavity is fixed by the boundary  to a low value for $R>\xi_{\mathrm{ps},\bm{x}}(T;\bm{r}_0^N)$ ($\Delta F_{\bm{x}}<0$) when the configurational entropy dominates the free energy. One recovers Eq.~(\ref{eqn:pts}), with $\xi_\mathrm{ps}(T)$, $\Sigma(T)$ and $\Upsilon(T)$ now replaced by the fluctuating quantities $\xi_{\mathrm{ps},\bm{x}}(T;\bm{r}_0^N)$, $\Sigma_{\bm{x}}^{(R)}(T;\bm{r}_0^N)$, and $\Upsilon_{\bm{x}}^{(R)}(T;\bm{r}_0^N)$ with $R=\xi_{\mathrm{ps},\bm{x}}(T;\bm{r}_0^N)$. The above argument predicts that the spatial fluctuations of the configurational entropy density and of the surface tension between amorphous states induce fluctuations of the point-to-set length which is determined through the self-consistent equation
\begin{equation}
\begin{aligned}
0=& -T\Sigma_{\bm{x}}^{(\xi_{\mathrm{ps},\bm{x}}(T;\bm{r}_0^N))}(T;\bm{r}_0^N)+\\&
\frac d{\rho [\xi_{\mathrm{ps},\bm{x}}(T;\bm{r}_0^N)]^{d-\theta}}\Upsilon_{\bm{x}}^{(\xi_{\mathrm{ps},\bm{x}}(T;\bm{r}_0^N))}(T;\bm{r}_0^N). 
\end{aligned}
\end{equation}
In particular, by varying the position $\bm{x}$ of the cavity and measuring the local point-to-set length~\cite{hocky2014correlation, charbonneau2016linking}, one should be able to describe some aspects of the self-induced disorder in glass-forming liquids.

However, this raises conceptual and technical issues. On the one hand, just from the spatial fluctuations of the point-to-set length it is impossible to disentangle the fluctuations of the configurational entropy from those of the surface tension. On the other hand, numerical measurements of the point-to-set length are challenging because small cavities are much slower to thermalize than their bulk counterpart~\cite{berthier2016efficient} and the simulations require enhanced sampling techniques such as parallel tempering~\cite{hukushima1996exchange}. Despite these limitations, the analysis of the average overlap profile in cavities with frozen boundaries has been shown to be consistent with a Weibull-distributed surface tension~\cite{cavagna2007mosaic, biroli2008thermodynamic}. 

\subsection{Mesoscopic systems with periodic boundary conditions}
\label{sec:stat_mech_bulk_eps}

Another way of probing the mesoscopic fluctuations of the configurational entropy is to consider relatively small systems of linear size $L$ with periodic boundary conditions and to bias its overlap with a reference configuration $\bm{r}_0^N$ with a linear field $\epsilon$, see Fig.~\ref{fig:settings}(c). This program was recently proposed in Ref.~[\onlinecite{PhysRevLett.127.088002}]. The free energy of the system now results from the competition between the configurational entropy contribution $T\Sigma^{(L)}(T;\bm{r}_0^N) \rho V_d L^d$ and an energy $-\epsilon\rho V_d L^d Q $ which attracts the system toward the reference configuration; $\Sigma^{(L)}(T;\bm{r}_0^N)$ is now the coarse-grained configurational entropy measured over the simulation box. Then, the free energy cost per unit particle associated with the system having a large overlap with the reference configuration is simply
\begin{equation}
  \Delta F(L,T;\bm{r}_0^N)=T\Sigma^{(L)}(T;\bm{r}_0^N)-\epsilon,
  \label{eq:pbc}
\end{equation}
where for simplicity we have assumed in the last term that the overlap is a binary variable, $Q=0,1$. The configurational entropy term is positive as it tends to push the system away from the reference configuration and it should be overcome by the field $\epsilon$ which instead attracts the system toward the reference configuration.

In this construction, the applied field $\epsilon$ is the control parameter. It induces a crossover from a low overlap when the configurational entropy dominates the free energy ($\Delta F>0$) to a high overlap when the attractive energy wins ($\Delta F<0$) at a value $\epsilon=\epsilon_{L}^*(T;\bm{r}_0^N)$, with 
\begin{equation}
\epsilon_{L}^*(T;\bm{r}_0^N)=T\Sigma^{(L)}(T;\bm{r}_0^N),
\label{eqn:eps_bulk}
\end{equation} 
as sketched in Fig.~\ref{fig:settings}(d).

Interestingly, due to the periodic boundary conditions, the surface tension does not appear in the free energy in Eq.~(\ref{eq:pbc}) and one can directly access the fluctuations of the configurational entropy density from those of the crossover applied field $\epsilon_{L}^*(T;\bm{r}_0^N)$. However, from a conceptual point of view, this method does not permit to reconstruct a spatially varying field of the configurational entropy density in a bulk macroscopic system and the boundary conditions are not well controlled, as the mesoscopic system interacts with itself. This geometry is however well suited to probe correlations with the local relaxation dynamics~\cite{PhysRevLett.127.088002}. From a practical point of view, there is a lower bound on the system size $L$: for too small values of $L$ the system with periodic boundary conditions tends to crystallize more easily~\cite{brumer2004numerical,PhysRevLett.127.088002}. 

\subsection{Local measurement of the Franz-Parisi potential with unconstrained boundaries}
\label{sec:stat_mech_local_eps}

We introduce a third geometry where we bias the overlap with a reference configuration via an applied field $\epsilon$ which only acts inside a mesoscopic cavity of radius $R$ around a position $\bm{x}$. Outside the cavity, the system freely evolves at thermal equilibrium with no thermodynamic constraint, see Fig.~\ref{fig:settings}(e). Compared to the two previous settings, this amounts to having a small overlap at the boundary of the cavity, and the field $\epsilon$ is then used to probe the overlap fluctuations inside the mesoscopic 
system. In this case, the free energy associated with keeping the cavity close to the reference configuration is given by
\begin{equation}
\Delta F_{\bm{x}}(R,T,\epsilon;\bm{r}_0^N)=T\Sigma_{\bm{x}}^{(R)}(T;\bm{r}_0^N)+\frac{d}{\rho R^{d-\theta}}\Upsilon_{\bm{x}}^{(R)}(T;\bm{r}_0^N)-\epsilon
\label{eqn:free_energy}
\end{equation}
where we have again assumed for simplicity that the overlap takes only two values, $0$ and $1$.

The above free energy $\Delta F_{\bm{x}}(R,T,\epsilon;\bm{r}_0^N)$ results from the competition between the configurational entropy which leads the cavity to explore the configuration space and the coupling $\epsilon$ which instead forces the cavity to remain close to the reference configuration, and one further needs to take into account the surface cost when the density profiles inside and outside the cavity have a mismatch. In this setting, the control parameter is still the applied field $\epsilon$. At a fixed cavity size $R$, the overlap in the cavity crosses over from a low value for $\epsilon<\epsilon_{\bm{x}}^*(R,T;\bm{r}_0^N)$ ($\Delta F_{\bm{x}}>0$) to a high value for $\epsilon>\epsilon_{\bm{x}}^*(R,T;\bm{r}_0^N)$ ($\Delta F_{\bm{x}}<0$) with
\begin{equation}
\epsilon_{\bm{x}}^*(R,T;\bm{r}_0^N)=T\Sigma^{(R)}_{\bm{x}}(T;\bm{r}_0^N)+\frac{d}{\rho R^{d-\theta}}\Upsilon_{\bm{x}}^{(R)}(T;\bm{r}_0^N),
\label{eqn:eps_star}
\end{equation}
as sketched in Fig.~\ref{fig:settings}(f). By varying the size $R$ of the cavity, one can in principle access both the configurational entropy density $\Sigma^{(R)}_{\bm{x}}(T;\bm{r}_0^N)$ and the local surface tension $\Upsilon_{\bm{x}}^{(R)}(T;\bm{r}_0^N)$. Varying the position ${\bm x}$ of the cavity then reveals the self-induced heterogeneity in the reference configuration and the spatial fluctuations of both $\Sigma_{\bm{x}}^{(R)}$ and $\Upsilon_{\bm{x}}^{(R)}$, thus allowing in principle the reconstruction of the spatial fields of these two quantities. 

One of the main advantages of this scheme is that the fluctuations of the configurational entropy and of the surface tension can be studied independently by varying $R$. This gives us a handle on the distribution of the emergent random-field and random-bond disorders. It should also be possible to extract the typical correlation length $\xi_\Sigma(T)$ of the configurational entropy that we have introduced above. This length characterizes the spatial extent of the correlations in the effective random field. To our knowledge, it has not been previously discussed in the literature, and its relation to the point-to-set length $\xi_\mathrm{ps}(T)$ is not known. From a practical point of view, this setting is also much less demanding in terms of computer resources than point-to-set measurements because the cavity can more easily thermalize in the absence of frozen constraints at its boundary; the outside of the cavity now acts as a reservoir of particles with unconstrained dynamics. 

\section{Numerical model and computational methods}
\label{sec:methods}

\subsection{Numerical model and equilibration}

We simulate a well-known size-polydisperse soft-sphere system of $N=576$ particles of equal mass $m$ in $d=2$ with a distribution of diameters $\mu(\sigma_i)\sim\sigma_i^{-3}$~\cite{ninarello2017models, berthier2019zero, guiselin2021microscopic, guiselin2021statistical} for $\sigma_i \in [\sigma_\mathrm{min},\ \sigma_\mathrm{max}]$ with $\sigma_\mathrm{max}/\sigma_\mathrm{min}\approx 2.225$. Two particles $i$ and $j$ interact via the repulsive potential $v(r)=v_0(\sigma_{ij}/r_{ij})^{12}+c_0+c_2(r_{ij}/\sigma_{ij})^2+c_4(r_{ij}/\sigma_{ij})^4$ if their relative distance $r_{ij}=|\bm{r}_i-\bm{r}_j|$ satisfies $r_{ij}/\sigma_{ij}<x_\mathrm{c}=1.25$. The constants $c_0$, $c_2$ and $c_4$ are chosen so that the potential and its two first derivatives are continuous at the cutoff $x_\mathrm{c}$: $c_0=-28 v_0/{x_\mathrm{c}}^{12}$, $c_2=48 v_0/{x_\mathrm{c}}^{14}$, $c_4=-21 v_0/{x_\mathrm{c}}^{16}$. This choice of $\mu(\sigma_i)$ and of nonadditive cross-diameters $\sigma_{ij}=(1-\eta|\sigma_i-\sigma_j|)(\sigma_i+\sigma_j)/2$ prevents crystallization and fractionation~\cite{ninarello2017models}. Energies and temperatures are expressed in units of $v_0$ (the Boltzmann constant is set to unity), lengthscales in units of the average diameter $\sigma$ of the particles and timescales in units of $\sqrt{m\sigma^2/v_0}$. Using these units, we set $\eta=0.2$, $\sigma_\mathrm{min} \approx 0.725$ and $\sigma_\mathrm{max}=1.613048$. The number density $\rho=N/L^d$ equals $1$, or, equivalently, we set the linear size $L$ of the system to $L=24$. This system has already been well characterized and we report here three conventional temperature scales~\cite{berthier2019zero, guiselin2021microscopic}: the onset temperature of glassy behavior $T_\mathrm{on}=0.2$, the mode-coupling crossover temperature $T_\mathrm{mct}=0.115$, and the extrapolated calorimetric glass transition temperature $T_\mathrm{g}=0.068$.

We first generate equilibrium configurations used for the reference configurations $\bm{r}_0^N$ at a temperature $T$ with the Hamiltonian $\widehat H[\bm{r}_0^N]=\sum_{i<j}v(r_{ij})$, where the sum runs over all pairs of particles with $i,j=1\dots N$. We use a hybrid scheme combining molecular dynamics (MD) with a Nos\'e-Hoover thermostat~\cite{nose1984unified, hoover1985canonical, martyna1992nose} and swap Monte Carlo moves of particle diameters that have been shown to drastically speedup equilibration~\cite{ninarello2017models, berthier2019efficient}. The equations of motion in the MD are solved with a time step $\mathrm{d}t=0.005$ and a thermostat damping time $\tau_\mathrm{th}=0.5$ by using a Liouvillian-based reversible integrator~\cite{martyna1996explicit, frenkel2001understanding}. After $n_\mathrm{MD}=50$ MD steps, the positions and the velocities of the particles are kept fixed and $N_\mathrm{swap}=n_\mathrm{swap}N$ swap Monte Carlo moves are attempted, with $n_\mathrm{swap}=10$. These two steps are then repeated and independent configurations are stored every 2$\tau_\alpha^\mathrm{swap}(T)$ where $\tau_\alpha^\mathrm{swap}$ represents the structural relaxation for the hybrid MD dynamics with swap moves. 

\subsection{Local measurement of the Franz-Parisi potential}

We simulate a second configuration $\bm{r}^N$ of the system at the same temperature $T$ using the same scheme but with the modified Hamiltonian  (with $d=2$) 
\begin{equation}
\widehat H_{\bm{x},\epsilon}^{(R)}[\bm{r}^N;\bm{r}_0^N]=\widehat H[\bm{r}^N]-\rho R^dV_d\epsilon\widehat Q_{\bm{x}}^{(R)}[\bm{r}^N;\bm{r}_0^N],
\end{equation}
following the local construction described in Sec.~\ref{sec:stat_mech_local_eps}. In the above expression, the local overlap in the cavity of radius $R$ at position $\bm{x}$ is 
defined as
\begin{equation}
\widehat Q_{\bm{x}}^{(R)}[\bm{r};\bm{r}_0^N]=\frac{\sum_{i,j=1}^Nw(|\bm{r}_i-\bm{r}_{j,0}|/a)\,\phi(|\bm{r}_i-\bm{x}|/R)}{\sum_{i=1}^N\phi(|\bm{r}_i-\bm{x}|/R)},
\label{eqn:ov}
\end{equation}
with $w(x)$ and $\phi(x)$ \rev{smooth versions of the Heaviside step function $\Theta(1-x)$ to avoid discontinuities in the forces exerted on the particles in the course of the molecular dynamics simulations.} For convenience, we choose $w(x)=e^{-x^4\ln 2}$ and $\phi(x)=\Theta(1-x)+\Theta(x-1)\Theta(1+b-x)\times[(e^{-(x-1)^4}-e^{-b^4})/(1-e^{-b^4}) + \kappa_2(x-1)^2+\kappa_3(x-1)^3]$, where the constants $\kappa_2$ and $\kappa_3$ enforce that $\phi(x)$ and its first derivative are continuous at $x=1$ and $x=1+b$, namely, $\kappa_2=-b\kappa_3$ and $\kappa_3=4b e^{-b^4}/(1-e^{-b^4})$. These specific expressions for $w(x)$ and $\phi(x)$ are expected to affect the results only quantitatively, leaving unchanged the qualitative trends presented in this work. They involve two lengthscales. The choice $a=0.22$ used in the definition of the overlap has been discussed before~\cite{guiselin2020overlap}, while $b=0.07 \ll R$ is a very small length defining a boundary layer for the mesoscopic cavity of radius $R$, \rev{thus mimicking the behavior of the Heaviside function.}

For a given reference configuration, given position and radius of the cavity, we simulate $n_\mathrm{s}\approx 10$ different values of $\epsilon$ in the range $[\epsilon_\mathrm{min},\ \epsilon_\mathrm{max}]$ and we monitor the histogram of overlap values in the cavity. The values of $\epsilon$ are chosen so that the different histograms significantly overlap and fill the entire range $\widehat Q_{\bm{x}}^{(R)}\in[0,1]$. We subsequently use histogram reweighting techniques to compute the probability distribution of the overlap for any field $\epsilon\in[\epsilon_\mathrm{min},\ \epsilon_\mathrm{max}]$~\cite{ferrenberg1989optimized, newman1999monte, kumar1992weighted}, and we eventually define the crossover field $\epsilon_{\bm{x}}^*$ as the field value for which the variance of the overlap in the cavity is maximum~\cite{guiselin2021statistical}. As in previous studies of overlap fluctuations~\cite{berthier2013overlap, guiselin2020random, guiselin2021statistical} we have carefully checked that all distributions are correctly sampled in fully equilibrium conditions. When dealing with ensemble averages we then consider a number $n_\mathrm{m}$ of independent samples, with $n_\mathrm{m} \in [3,\ 13]$ depending on the temperature.

The radius $R$ should be taken small enough, ideally smaller than the correlation length of the configurational entropy to resolve its intrinsic fluctuations and smaller than the point-to-set length; but it should be sufficiently large so that the cavity contains enough particles to compute well-defined mesoscopic quantities, such as the overlap in Eq.~(\ref{eqn:ov}). 
In the temperature range investigated here, $\xi_\mathrm{ps}$ is at most equal to $4$~\cite{berthier2019zero}. We thus focus on $R=2$ and $R=4$ (while the linear system size is $L=24$), respectively corresponding to 13 and 52 particles on average in a cavity.

For a given reference configuration, we consider cavity centers on a linear grid of mesh size $u=2$, corresponding to $(L/u)^d=144$ different positions, and for each of them we compute the crossover field $\epsilon_{\bm{x}}^*(R,T;\bm{r}_0^N)$. Then, a continuous and coarse-grained field ${\overline \epsilon}_{\bm{x}}(R,T;\bm{r}_0^N)$ is computed by using a Gaussian window of width $\ell=u/2=1$ and by summing over all the positions of the cavities, namely, 
\begin{equation}
{\overline \epsilon}_{\bm{x}}(R,T;\bm{r}_0^N)=\frac{\sum_{\bm{x}'}\epsilon_{\bm{x}'}^*(R,T;\bm{r}_0^N)e^{-|\bm{x}-\bm{x}'|^2/(2\ell^2)}}{\sum_{\bm{x'}}e^{-|\bm{x}-\bm{x}'|^2/(2\ell^2)}}.
\label{eqn:cg_eps}
\end{equation}
This allows us to associate to each particle $i$ a local field ${\overline \epsilon}_i(R,T;\bm{r}_0^N)$ obtained from the above equation with $\bm{x}=\bm{r}_i$ and resulting from a coarse-grained computation. 

This work represents a significant computational effort: each configuration requires the independent study of a large number of cavities, and each cavity requires itself a series of lengthy simulations. The additional ensemble average then multiplies the needed effort, which must be repeated for each temperature. Although the study is trivially parallelized, it requires 
a large amount of numerical resources. 

\section{Results}
\label{sec:results}

We are now in a position to measure the local fluctuations of the overlap and access the static spatial variations of the Franz-Parisi potential and of the configurational entropy density across a broad range of temperatures.  

\subsection{Spatial maps of the crossover field in cavities}

\begin{figure}[t]
\includegraphics[width=0.49\linewidth]{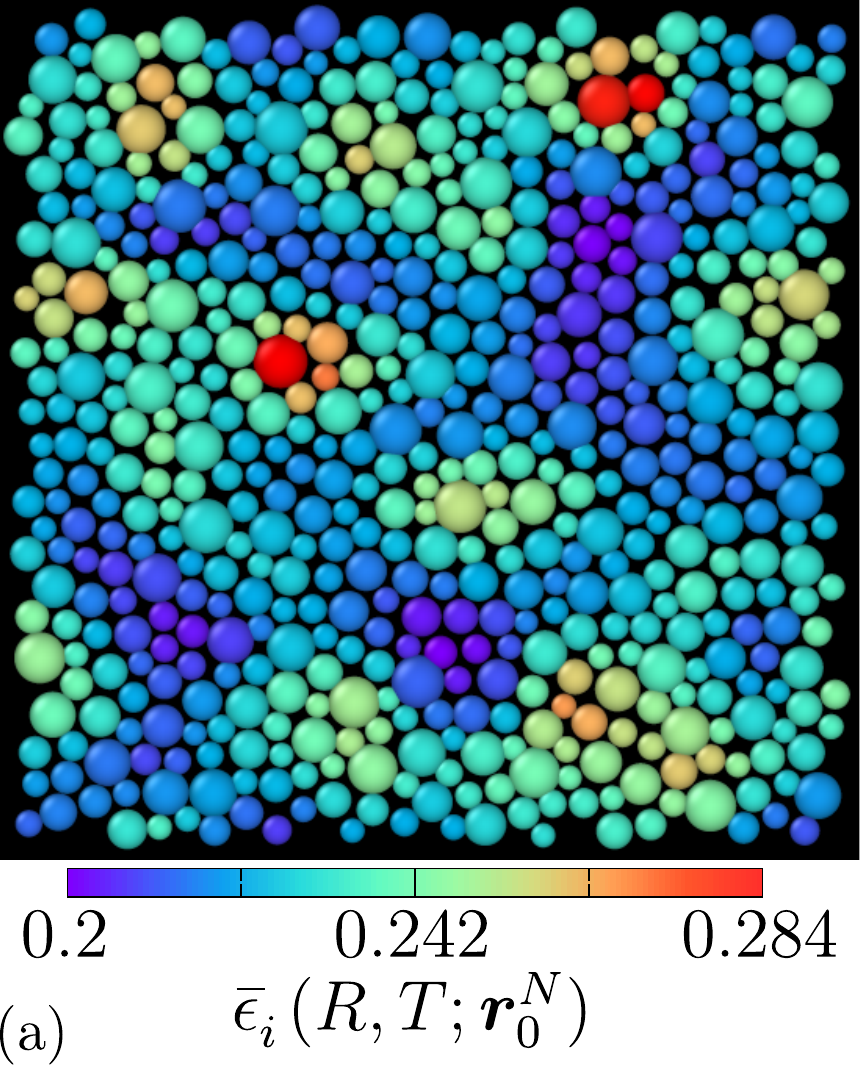}
\includegraphics[width=0.49\linewidth]{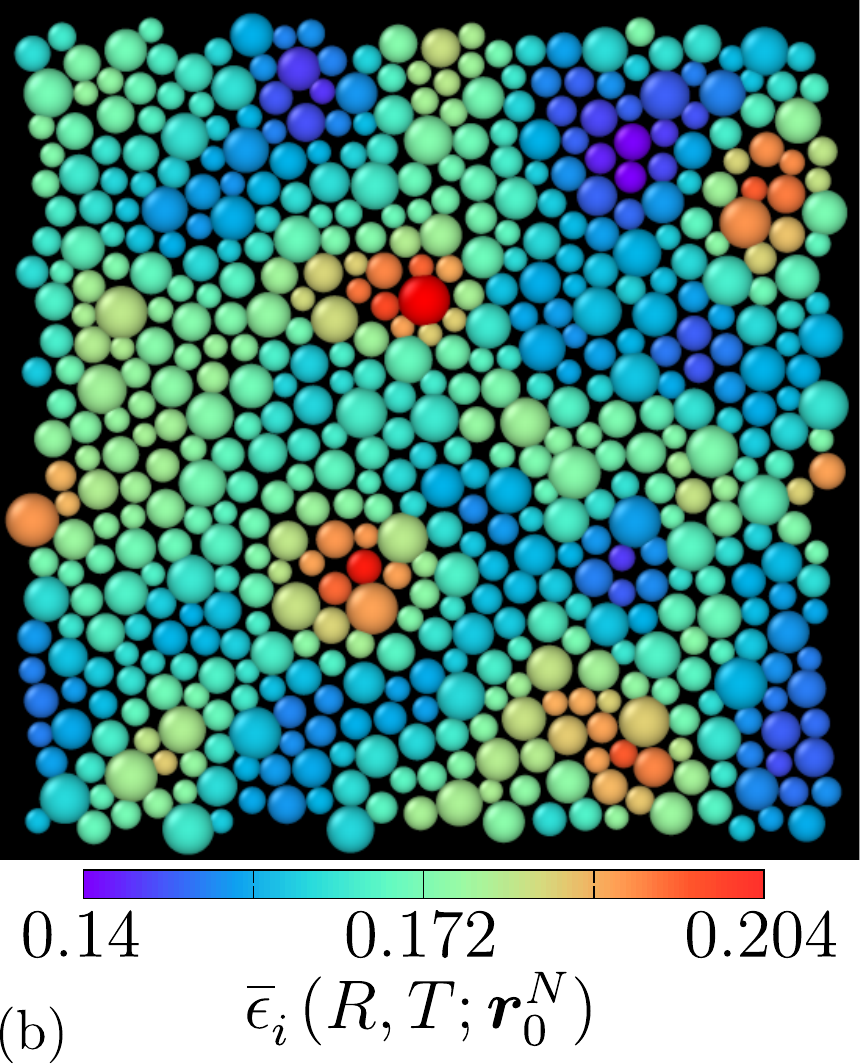}
\includegraphics[width=0.49\linewidth]{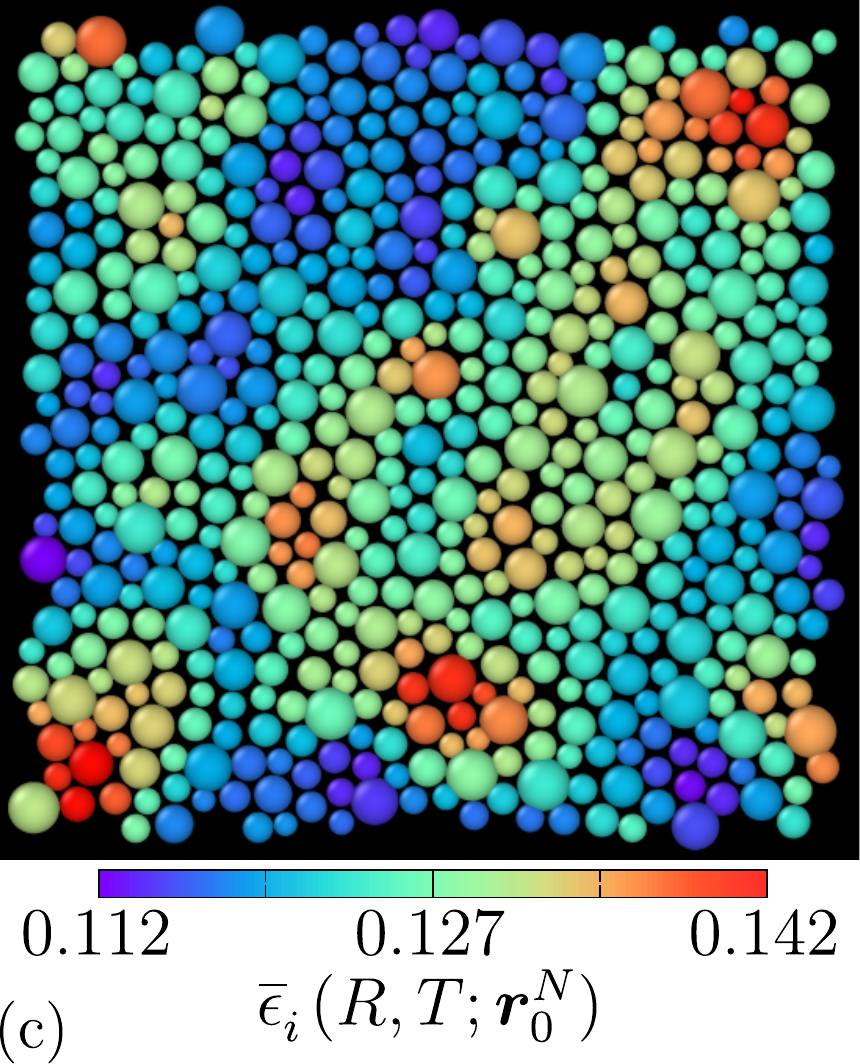}
\includegraphics[width=0.49\linewidth]{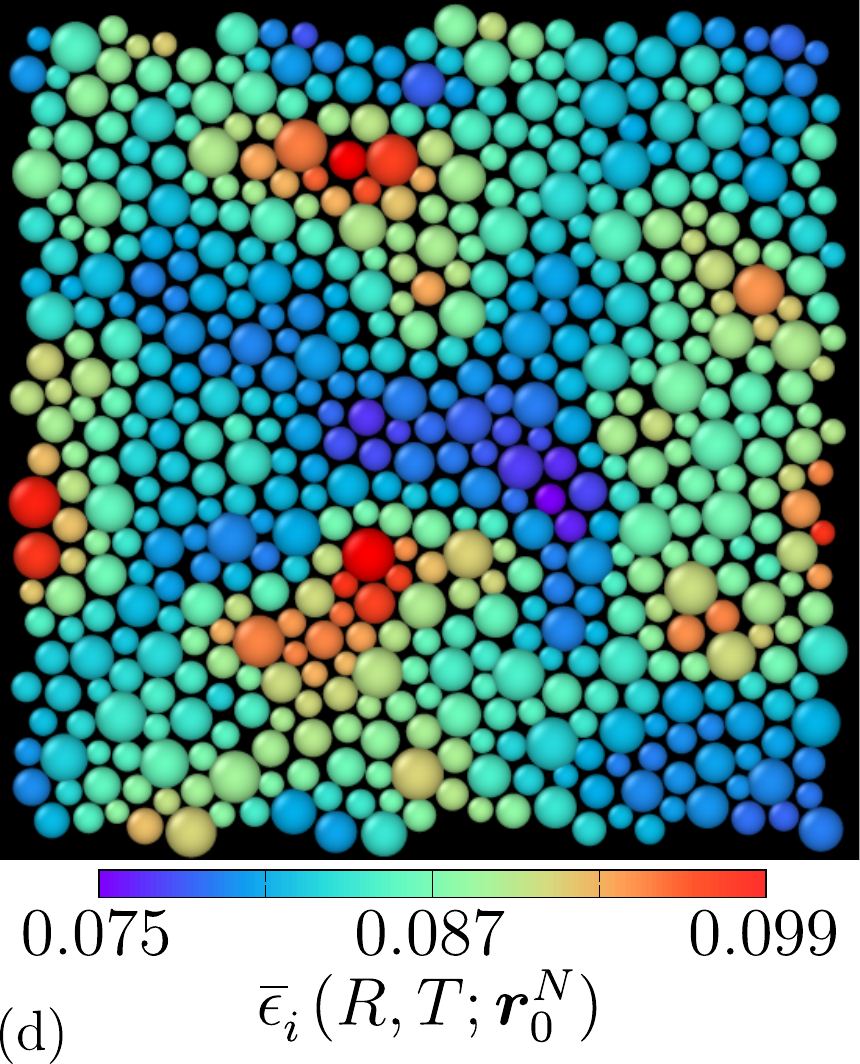}
\caption{Map of the coarse-grained crossover field ${\overline \epsilon}_i(R,T;\bm{r}_0^N)$ for $R=2$ in a system with $L=24$ and an equilibrium reference configuration 
$\bm{r}_0^N$ sampled at different temperatures $T$: (a) $T=0.165$, (b) $T=0.125$, (c) $T=0.1$ and (d) $T=0.07$. The color code is adjusted independently for each panel.}
\label{fig:map}
\end{figure}

The basic outcome of the simulations described in Sec.~\ref{sec:methods} is the evolution of the local overlap at position ${\bm x}$ with a local field also applied at ${\bm x}$ throughout the entire system. From these overlap isotherms, we extract the crossover applied field, which then depends on space for each specific reference configuration. This can be repeated for independent reference configurations at various temperatures. 

Using the coarse-graining procedure in Eq.~(\ref{eqn:cg_eps}), we construct maps representing the crossover field ${\overline \epsilon}_i(R,T;\bm{r}_0^N)$ attributed to each particle. In Fig.~\ref{fig:map} we show representative snapshots of this crossover field for $R=2$ and several reference configurations at temperatures covering a very broad range from much above the mode-coupling crossover down to very close to the calorimetric glass transition temperature. 

These snapshots reveal that the field ${\overline \epsilon}_{i}$ fluctuates widely within a given configuration. Because this crossover field represents an estimate of the configurational entropy density, these images directly show that the configurational entropy density in an equilibrium supercooled liquid is a spatially-fluctuating quantity. These measurements represent a direct and quantitative visualization of the physical concept of the self-induced disorder characterizing glass-forming liquids~\cite{bouchaud1994self}.

Despite their apparent visual similarity, notice that the color scale has been independently adjusted in each snaphot to maximize the color contrast, and both the average level and spread of the field are actually temperature dependent. In the following we quantify the field fluctuations in detail.

Still, the images do not appear to display a strong evolution with temperature of the typical spatial extent of the fluctuations. For this reason, we have not attempted a more precise characterization by using a spatial correlation function. Since the field is already coarse-grained over a domain of diameter $2R=4$ and the point-to-set length is at most $\xi_{\rm ps}\approx 4$ in the temperature range under investigation~\cite{berthier2019zero} (and one may anticipate that $\xi_{\rm ps}$ is an upper bound for all thermodynamic correlation lengths), we do not expect any detectable variation of the correlation length $\xi_{\Sigma}$ in this range. An interesting goal for future work would be to study a temperature regime in which the point-to-set length becomes much larger than the coarse-graining length, which should be possible via the swap Monte Carlo algorithms~\cite{ninarello2017models, berthier2019efficient}. This would allow one to determine whether $\xi_{\Sigma}(T)$ eventually decouples from $\xi_{\rm ps}(T)$.

\subsection{Average configurational entropy and surface tension}

\begin{figure}[t]
\includegraphics[width=0.99\columnwidth]{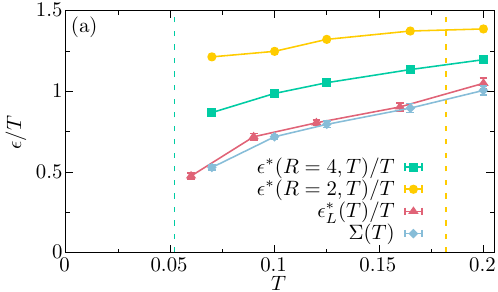}
\includegraphics[width=0.99\columnwidth]{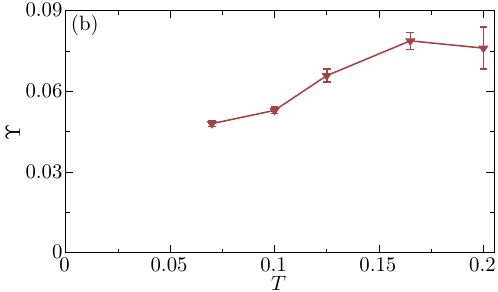}
\caption{(a) Average crossover field $\epsilon^*(R,T)=\langle \epsilon^*_{\bm{x}}(R,T;\bm{r}_0^N) \rangle$ for cavity sizes $R=2,\ 4$ along with the average $\epsilon^*_{L}(T)$ in bulk systems of linear size $L=8$. All fields are rescaled by the temperature $T$. We also represent the estimate of the configurational entropy $\Sigma(T)$ obtained from Eq.~(\ref{eqn:sigma}). The vertical dashed lines mark $T=0.182$ and $T=0.052$ for which the point-to-set length equals $R=2$ and $R=4$, respectively. (b) Evolution of the average surface tension from  Eq.~(\ref{eqn:upsilon}). All error bars are computed by the jackknife method when averaging over reference configurations and are not shown when smaller than the symbols.}
\label{fig:avg}
\end{figure}

In order to have more insight into the statistics of the spatial fluctuations of the crossover field shown in Fig.~\ref{fig:map}, we first consider its average $\epsilon^*(R,T)= \langle \epsilon^*_{\bm{x}}(R,T;\bm{r}_0^N) \rangle$ over reference configurations. \rev{We find that $\epsilon^*(R,T)$ decreases rapidly  with decreasing temperature. This is consistent with the fact that the attraction between replicas competes with a driving force of entropic nature, as captured by Eq.~(\ref{eqn:eps_star}). It is therefore more convenient to display} the temperature evolution of this average rescaled by the temperature $T$ for $R=2$ and $R=4$ in Fig.~\ref{fig:avg}(a).

We observe that the rescaled crossover field decreases with decreasing temperature as could be anticipated from the direct inspection of the maps in Fig.~\ref{fig:map}. This is expected if the evolution of $\epsilon^*/T$ is dominated by that of the configurational entropy density. The variation of the latter with the temperature has already been reported for this system~\cite{berthier2019zero} and indeed decreases as $T$ decreases.

The average crossover field $\epsilon^*(R,T)$ also shifts toward smaller values when the radius $R$ of the cavity increases at fixed temperature $T$. It always lies above its counterpart $\epsilon_{L}^*(T)$ measured in system of linear size $L=8$ ($=2R$ for $R=4$) with periodic boundary conditions (see Ref.~[\onlinecite{guiselin2021statistical}] for the methods). 
\rev{These observations are qualitatively consistent with the fact that an extra free-energy cost due to the mismatch between the density profiles inside and outside the cavity increases the value of the crossover field. This extra contribution increases when $R$ decreases, suggesting that it does not scale with the volume $R^d$ but indeed behaves as a surface tension term.}

We rationalize the behavior of $\epsilon^*(R,T)/T$ seen in Fig.~\ref{fig:avg}(a) as follows. When $T$ is reduced, the configurational entropy decreases, possibly to zero, while the surface tension is expected to remain finite~\cite{franz2005first}. The competition between these two contributions to the free energy is precisely ruled by the growth of the point-to-set length, see Eq.~(\ref{eqn:eps_star}). From Eq.~(\ref{eqn:eps_star}) one then roughly predicts that the configurational entropy controls the evolution of $\epsilon^*(R,T)/T$ at high temperature, while the surface tension dominates at low temperature. The crossover between the high- and low-temperature regimes is expected around the temperature for which the radius $R$ of the cavity is of the order of the point-to-set length. As seen in Fig.~\ref{fig:avg}(a), $\epsilon^*(R=4,T)/T$ indeed roughly follows the evolution of $\epsilon_{L}^*(T)/T$ (which is controlled by the configurational entropy density only) down to a temperature approaching that at which $\xi_{\rm ps}\approx 4$ whereas $\epsilon^*(R=2,T)/T$ deviates already from $\epsilon_{L}^*(T)/T$ at the highest temperatures where $\xi_{\rm ps}\gtrsim 2$.

We can go one step further and assume that Eq.~(\ref{eqn:eps_star}), which qualitatively accounts for our observations, is in fact quantitatively valid. If correct, then 
$\epsilon^*(R,T)=T\Sigma(T)+d\Upsilon(T)/(\rho R^{d-\theta})$; $\Upsilon(T)$ here stands for the average surface tension in the case of a spherical interface and is assumed to be independent of $R$, which neglects the curvature effects that may be present at small $R$~\cite{tolman1949effect} and the possible random-field-like interface behavior at large $R$~\cite{grinstein1983surface}. Under these conditions one can extract the average configurational entropy as
\begin{equation}
\Sigma(T)=\frac{2^{d-\theta}\epsilon^*(R,T)-\epsilon^*(R/2,T)}{T\left(2^{d-\theta}-1\right)},
\label{eqn:sigma}
\end{equation}
with $R=4$, $d=2$, and $\theta=d-1=d/2=1$. The application of Eq.~(\ref{eqn:sigma}) is shown in Fig.~\ref{fig:avg}(a) and agrees very well with the direct measurement of $\epsilon_L^*(T)/T$ in bulk systems, for which the surface tension plays no role [recall Eq.~(\ref{eqn:eps_bulk})]. This agreement supports the validity of Eq.~(\ref{eqn:eps_star}).

One can similarly extract the average surface tension,
\begin{equation}
\Upsilon(T)=\frac{\rho R^{d-\theta}\left[\epsilon^*(R/2,T)-\epsilon^*(R,T)\right]}{d\left(2^{d-\theta}-1\right)},
\label{eqn:upsilon}
\end{equation}
with $R=4$, $d=2$ and $\theta=1$. The result is shown in Fig.~\ref{fig:avg}(b) as a function of the temperature $T$. The average surface tension decreases by $\sim 40\%$ 
when the temperature decreases, while at the same time $\Upsilon(T)/T$ is found to grow. Past works dealing with the surface tension have reached contradictory conclusions regarding its temperature evolution. In Refs.~[\onlinecite{cammarota2009evidence}, \onlinecite{cammarota2009numerical}], the surface tension was found to increase with decreasing temperature in agreement with instanton calculations~\cite{franz2005first, dzero2005activated}. Instead, the surface tension is usually taken proportional to the temperature, as for conventional phase separation problems, in many analyses performed in the context of the RFOT theory~\cite{xia2000fragilities, lubchenko2007theory}, suggesting that it is a increasing function of the temperature. Our simulation data appear intermediate between these two proposals. 

We can tentatively rationalize the variation of the surface tension with temperature in Fig.~\ref{fig:avg}(b) by recalling that our estimate may differ from the actual surface tension between states because of the possibly more complex geometry of the RFOT mosaic. If the domains composing the mosaic become more compact at lower temperatures~\cite{stevenson2006shapes, biroli2017fluctuations}, our method could overestimate the surface tension at high temperatures by neglecting these geometrical effects. 

\subsection{Fluctuations of the configurational entropy: variance and correlation length}

\begin{figure}[t]
\includegraphics[width=0.99\columnwidth]{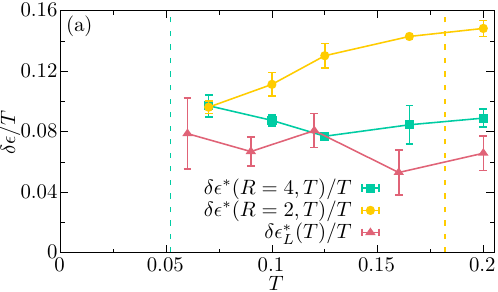}
\includegraphics[width=0.99\columnwidth]{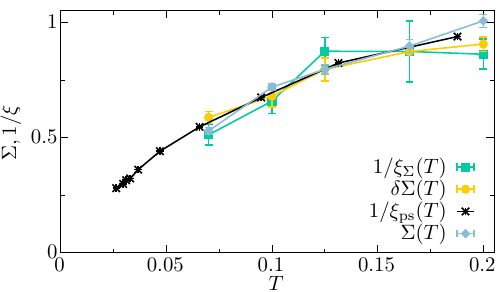}
\caption{(a) Standard deviation $\delta\epsilon^*(R,T)$ of the crossover field for cavity sizes $R=2,\ 4$, and for a bulk system of linear size $L=8$. The vertical dashed lines mark $T=0.182$ and $T=0.052$ for which the point-to-set length is equal to $2$ and $4$, respectively. (b) Evolution of the inverse correlation length $1/\xi_\Sigma(T)$ from Eq.~(\ref{eqn:xi_sigma}), of the standard deviation $\delta\epsilon^*(R=2,T)/T = \delta\Sigma(T)$, of the inverse point-to-set length $1/\xi_\mathrm{ps}(T)$ from Ref.~[\onlinecite{berthier2019zero}], and of the averaged configurational entropy $\Sigma(T)$ from Eq.~(\ref{eqn:sigma}). The first three curves have been rescaled by a constant to maximize their overlap.}
\label{fig:std}
\end{figure}

We now go beyond the average behavior and analyze the fluctuations of the crossover field quantitatively. In Fig.~\ref{fig:std}(a), we display the temperature evolution of the standard deviation,
\begin{equation}
\delta\epsilon^*(R,T)= \sqrt{\langle [\epsilon^*_{\bm{x}}(R,T;\bm{r}_0^N)-\epsilon^*(R,T)]^2\rangle},
\end{equation}
rescaled by the temperature $T$ for the cavity sizes $R=2$ and $R=4$. We first observe that at fixed temperature, this standard deviation is larger for $R=2$ than for $R=4$. This is expected because the fluctuations are generically stronger in smaller systems. In addition, from Eq.~(\ref{eqn:eps_star}), the surface tension term also constitutes a stronger source of fluctuations for smaller $R$.

For comparison we also show in Fig.~\ref{fig:std}(a) the standard deviation
\begin{equation}
\delta \epsilon^*_{L}(T)= \sqrt{ \langle [\epsilon^*_{L}(T;\bm{r}_0^N)-\epsilon_{L}^*(T)]^2\rangle}
\end{equation}
of the crossover field measured in a bulk system with $L=8$. It is quite close to the results for $\delta \epsilon^*(R=4,T)$, especially at low temperature. This agreement suggests that the fluctuations of the configurational entropy dominate the fluctuations of $\epsilon^*_{\bm{x}}$ and that the surface tension contribution is subdominant. Therefore it is reasonable to assume that
\begin{equation}
\delta\epsilon^*(R,T) / T \approx \delta\Sigma^{(R)}(T),
\end{equation}
where $\delta\Sigma^{(R)}(T)$ represents the standard deviation of $\Sigma^{(R)}_{\bm{x}}$. This quantity should cross over from $\delta\Sigma(T)=\sqrt{\langle [\Sigma_{\bm{x}}(T;\bm{r}_0^N)-\Sigma(T)]^2\rangle}$ for $R<\xi_\Sigma(T)$, to $\delta\Sigma(T)[\xi_\Sigma(T)/R]^{d/2}$ for $R>\xi_\Sigma(T)$.

The temperature evolution of $\delta\epsilon^*(R,T)/T$ is more pronounced for $R=2$ (it decreases when $T$ decreases) than for $R=4$ (it is nearly constant), but both quantities seem to converge for $T\leq 0.07$. This difference in the temperature dependence is naturally explained if one assumes  that $\xi_\Sigma(T) \approx \xi_{\rm ps}(T)$ and grows with decreasing temperature from $\xi_\Sigma \approx 2$ to $\xi_\Sigma \approx 4$ in the range $T\in[0.07,\ 0.2]$~\cite{berthier2019zero}. This would indeed imply that for $R=2$, the system is always in the regime  $R<\xi_\Sigma(T)$, leading to $\delta \epsilon^*(R=2,T) / T= \delta \Sigma(T)$. Instead for $R=4$ one would explore the opposite regime $R > \xi_\Sigma(T)$ where $\delta\epsilon^* (R=4,T)/T = \delta \Sigma(T)\xi_\Sigma(T)/4$ (recall that $d=2$). If this assumption is correct, one can then combine these two expressions to obtain an estimate for the entropy correlation length valid for $T>0.07$:
\begin{equation}
\xi_\Sigma(T)=4  \frac{\delta \epsilon^*(R=4,T)}{\delta \epsilon^*(R=2,T)}. 
\label{eqn:xi_sigma}
\end{equation}

From the observation that $\delta\epsilon^*(R=4,T)/T = \delta \Sigma(T)\xi_\Sigma(T)/4$ is nearly constant, we deduce that $\delta \Sigma(T) \sim 1/\xi_\Sigma(T)$. Combining this with the above remark that $\xi_{\Sigma}\approx \xi_{\rm ps}$, we conclude that $\delta \Sigma(T) \sim 1/\xi_{\rm ps}(T) \sim \Sigma(T)$. This is in agreement with the RFOT theory which predicts that $\delta \Sigma(T)\sim [\Sigma(T)]^{d/[2(d-\theta)]} \sim \Sigma(T)$~\cite{dzero2009replica} in $d=2$ with $\theta=1$.

We test the self-consistency of this series of assumptions in Fig.~\ref{fig:std}(b) where we represent four different quantities: (i) $1/\xi_\Sigma(T)$ obtained by using Eq.~(\ref{eqn:xi_sigma}), (ii) $\delta\epsilon^*(R=2,T)/T$, (iii) $1/\xi_\mathrm{ps} (T)$ taken from Ref.~[\onlinecite{berthier2019zero}], and (iv) the configurational entropy $\Sigma(T)$ estimated by using Eq.~(\ref{eqn:sigma}). It can be seen that the four quantities evolve with temperature in essentially the same way. The quantities (i)-(iii) are rescaled by a constant factor, as they are defined up to an arbitrary prefactor. The very good agreement found between these four quantities confirms our hypothesis that the standard deviation for $R=2$ essentially follows the temperature evolution of the configurational entropy density itself, while its comparison with the result $R=4$ provides an estimate for the entropy correlation length which is in good agreement with the known evolution of the point-to-set-correlation length, $\xi_\Sigma(T) \sim \xi_{\rm ps}(T)$ in the studied temperature regime. \rev{As shown in Fig.~\ref{fig:std}, the configurational entropy and the point-to-set correlation length display a temperature evolution that is compatible with measurements performed in several glass-forming models~\cite{berthier2019configurational}. This modest temperature evolution has given rise to debates regarding the relevance of thermodynamic flucutations to account for the physics of supercooled liquids: see for instance Refs.\cite{PhysRevLett.119.195501,doi:10.1063/1.5086509}.}

Finally, these considerations about the correlation lengths rationalize the absence of a qualitative change in the maps of Fig.~\ref{fig:map}, as the point-to-set length only grows moderately in the temperature range investigated, and this modest evolution is masked by the coarse-graining procedure used to represent the random field.  

\subsection{Fluctuations of the configurational entropy: probability distribution}

\begin{figure}[t]
\includegraphics[width=0.99\columnwidth]{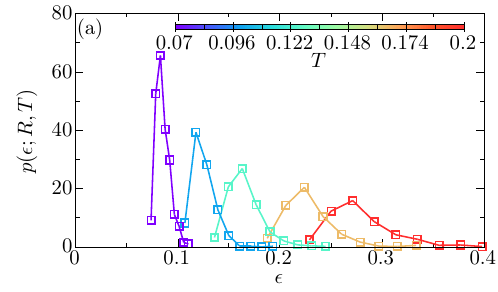}
\includegraphics[width=0.99\columnwidth]{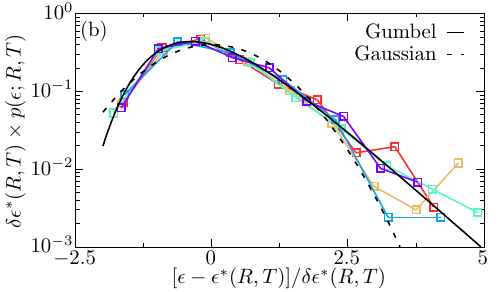}
\caption{Probability distribution $p(\epsilon;R,T)$ of the crossover field $\epsilon_{\bm{x}}^*(R=2,T;\bm{r}_0^N)$ for several temperatures $T$ as a function of (a) $\epsilon$ and (b) $X = [\epsilon-\epsilon^*(R,T)]/\delta\epsilon^*(R,T)$. The full line corresponds to a generalized Gumbel distribution with $\alpha \approx 1.5$, see Eq.~(\ref{eqn:gumbel}), while the dashed line is a Gaussian distribution.}
\label{fig:histo}
\end{figure}

We finally analyze the full probability distribution of $\epsilon^*_{\bm{x}}$ for $R=2$ and its temperature evolution, see Fig.~\ref{fig:histo}(a). In agreement with the snapshots shown in Fig.~\ref{fig:map}, the distribution narrows and shifts to lower values with decreasing temperature. Moreover, the distributions are all asymmetric, with a positive skewness, and 
display an extended tail toward large field values.

This is further confirmed by replotting the same data in Fig.~\ref{fig:histo}(b) with the help of the dimensionless variable
\begin{equation}
  X \equiv [\epsilon-\epsilon^*(R,T)]/\delta\epsilon^*(R,T),
\end{equation}
which is defined such that the average of $X$ is zero and its variance is unity. In this representation, all the data collapse on a single temperature-independent mastercurve $\mathcal{P}(X;R)$. Clearly, $\mathcal{P}(X;R)$ is not a Gaussian distribution. It decays slowly at large positive $X$ with an exponential tail and decays much more rapidly at large negative values.

Empirically, we find that the mastercurve is well fitted by a generalized Gumbel distribution with a single free parameter $\alpha \approx 1.5$~\cite{bramwell1998universality, bramwell2000universal, bramwell2001magnetic}:
\begin{equation}
\mathcal{P}(X;R)=\frac{\nu_\alpha \alpha^\alpha}{\Gamma(\alpha)} e^{\displaystyle -\alpha\left[\nu_\alpha(X+\lambda_\alpha)+e^{-\nu_\alpha(X+\lambda_\alpha)}\right]},
\label{eqn:gumbel}
\end{equation}
where $\Gamma(\alpha)$ is the Euler Gamma function and where the parameters $\nu_\alpha=\sqrt{(\ln\Gamma)''(\alpha)}$ and $\lambda_\alpha=\nu_\alpha^{-1}[\ln\alpha-(\ln\Gamma)'(\alpha)]$ involve the first two derivatives of the natural logarithm of $\Gamma(\alpha)$, denoted with primes. The distributions for $R=4$ can also be described by the same law but with a slightly smaller skewness (about 1.14 and 0.74 for $R=2$ and $R=4$, respectively). It would be interesting to repeat the analysis for even larger values of $R$ to check if the distributions $p(\epsilon;R,T)$ become Gaussian, as expected from the central-limit theorem when $R \gg \xi_\Sigma(T)$. 

At this stage, the origin of such a Gumbel distribution independent of the temperature remains somewhat unclear. The analogy with the results of Ref.~[\onlinecite{bramwell2000universal}] may originate from the fact that we record the statistics of the fluctuations of an observable defined over a mesoscopic length which is smaller than (or comparable to) its correlation length, and therefore appears `critical'. It would be valuable to repeat the analysis developed in this work with other model glass-formers in order to assess the universality of this distribution of crossover fields among dimension and models. In Ref.~[\onlinecite{PhysRevLett.127.088002}], a temperature-independent Gaussian distribution was obtained in relatively small three-dimensional hard-sphere systems with periodic boundary conditions, which either suggests that the distribution of the configurational entropy density is not universal, or that the linear size of the system in Ref.~[\onlinecite{PhysRevLett.127.088002}] was sufficiently large with respect to $\xi_\Sigma(T)$ for the central-limit theorem to hold. A final possibility is that surface-tension fluctuations present in our setting induce a quantitative difference between the probability distributions of the crossover field and those of the configurational entropy density even though they do not contribute much to the variance. 

\section{Conclusion and perspectives}
\label{sec:ccl}

We have introduced and numerically implemented a new probe to reveal the static self-induced disorder in glass-forming liquids. To this end, we have considered the statistics of the local overlap between pairs of liquid configurations within a mesoscopic cavity of linear size $R$ located at position ${\bm x}$. In particular, we have applied a field $\epsilon$ that is linearly coupled to the overlap inside the cavity, leaving the outside fully unconstrained. This geometry corresponds to a different setting than the usual point-to-set construction, in particular at the level of the boundary conditions. This is also conceptually different from the study of mesoscopic systems with periodic boundary conditions. 

Varying the location of the cavity in a systematic way allows us to scan the spatial fluctuations of the emergent disorder which is expected to take the form of random fields (configurational entropy density) and random bonds (local surface tension). We have indeed found nontrivial fluctuations in space of the crossover field $\epsilon^*_{\bm{x}}(R,T;\bm{r}_0^N)$ needed to induce a large local overlap inside the cavity. We have also obtained quantitative insight about the temperature evolution of the average surface tension. 

Although the crossover field $\epsilon^*_{\bm{x}}(R,T;\bm{r}_0^N)$ contains information concerning the local surface tension, the effect of the latter is rather small for the cavity sizes $R$ that we consider and in a first approximation the fluctuations of $\epsilon^*_{\bm{x}}(R,T;\bm{r}_0^N)$ can be taken as a proxy for those of the configurational entropy density. The maps in Fig.~\ref{fig:map} then represent a visualization of realizations of the effective random-field disorder associated with an equilibrium reference configuration. Such maps should not be confused with snapshots of the mosaic state predicted by the RFOT theory. The random-field disorder is an ingredient for an effective Hamiltonian describing glass-forming liquids at a coarse-grained level~\cite{biroli2018random1, biroli2018random2} (on a lengthscale of the order of $\xi_\Sigma$) \rev{whereas the mosaic state should in principle be obtained from the full statistical-mechanical treatment of this effective Hamiltonian.}
Of course, the characteristic lengthscales being rather limited in size, it is not always easy to disentangle the various levels in practice. One interesting piece of information would be to more systematically compare the correlation length of the effective self-induced disorder $\xi_\Sigma$ with the point-to-set length $\xi_{\rm ps}$. In the range of temperature that we have studied, we have found them roughly equal but this may change at lower temperature where one would anticipate a slower temperature variation for $\xi_\Sigma$ than for $\xi_{\rm ps}$.

\rev{From the viewpoint of the RFOT theory, our main contribution is the development of a concrete numerical method that can directly probe the existence and the spatial variations of well-defined analogs of the configurational entropy and of the surface tension between amorphous density profiles. Having measured the distribution of the effective random field coupled to the overlap for this model, one could now imagine building an effective field theory of the overlap in finite dimensions. 
This would be useful in order to address central questions posed by the application of the RFOT theory, in particular regarding the possible existence of a finite-temperature Kauzmann transition in three-dimensional glass-formers.}  

In future work, it would also be interesting to apply our framework to different model glass-formers and to perform a more systematic analysis of the fluctuations of the configurational entropy  density and of the surface tension. Extending our work to three-dimensional models is conceptually simple, and this should also be done in the future. It would also be useful to compare our approach to the more mechanistic~\cite{mizuno2013measuring, shakerpoor2020stability, widmer2008irreversible, barbot2018local, doi:10.1063/1.5024776} and geometric~\cite{coslovich2011locally, malins2013identification1, malins2013identification, tong2018revealing, paret2020assessing, boattini2020autonomously} studies of structural heterogeneity mentioned in the introduction. \rev{This would in particular illuminate the conceptual difference between the self-induced disorder considered here and the more mundane structural disorder characterizing aperiodic materials.}

Another natural direction of study would consist in assessing the connection between the self-induced static heterogeneity shown in this work and the well-known dynamic heterogeneity characterizing the structural relaxation of deeply supercooled liquids. The logarithm of the equilibrium relaxation time from different initial configurations was found to be positively correlated with the inverse of the crossover field or of the configurational entropy in mesoscopic bulk samples~\cite{PhysRevLett.127.088002, coslovich2016structure}, in direct agreement with the RFOT and Adam-Gibbs scenarios~\cite{bouchaud2004adam, ozawa2019does, adam1965temperature}. Analyzing whether a similar correlation holds at the mesoscopic scale may pave the way toward a better theoretical understanding of the complex dynamics of glass-forming materials which can now be numerically studied in deeply supercooled states~\cite{guiselin2021microscopic}. \rev{We consider this question as the most pressing task to assess the relevance of the RFOT theory description of glassy phenomena.}

\begin{acknowledgments}
We thank G. Biroli for useful discussions. Some simulations were performed at MESO@LR-Platform at the University of Montpellier. B. Guiselin acknowledges support by Capital Fund Management - Fondation pour la Recherche. This work was supported by a grant from the Simons Foundation (Grant No. 454933, L.B.).
\end{acknowledgments}

\section*{Data availability}

The data that support the findings of this study are available
from the corresponding author upon reasonable request.

\bibliography{biblio.bib}

\begin{thebibliography}{96}%
\makeatletter
\providecommand \@ifxundefined [1]{%
 \@ifx{#1\undefined}
}%
\providecommand \@ifnum [1]{%
 \ifnum #1\expandafter \@firstoftwo
 \else \expandafter \@secondoftwo
 \fi
}%
\providecommand \@ifx [1]{%
 \ifx #1\expandafter \@firstoftwo
 \else \expandafter \@secondoftwo
 \fi
}%
\providecommand \natexlab [1]{#1}%
\providecommand \enquote  [1]{``#1''}%
\providecommand \bibnamefont  [1]{#1}%
\providecommand \bibfnamefont [1]{#1}%
\providecommand \citenamefont [1]{#1}%
\providecommand \href@noop [0]{\@secondoftwo}%
\providecommand \href [0]{\begingroup \@sanitize@url \@href}%
\providecommand \@href[1]{\@@startlink{#1}\@@href}%
\providecommand \@@href[1]{\endgroup#1\@@endlink}%
\providecommand \@sanitize@url [0]{\catcode `\\12\catcode `\$12\catcode
  `\&12\catcode `\#12\catcode `\^12\catcode `\_12\catcode `\%12\relax}%
\providecommand \@@startlink[1]{}%
\providecommand \@@endlink[0]{}%
\providecommand \url  [0]{\begingroup\@sanitize@url \@url }%
\providecommand \@url [1]{\endgroup\@href {#1}{\urlprefix }}%
\providecommand \urlprefix  [0]{URL }%
\providecommand \Eprint [0]{\href }%
\providecommand \doibase [0]{http://dx.doi.org/}%
\providecommand \selectlanguage [0]{\@gobble}%
\providecommand \bibinfo  [0]{\@secondoftwo}%
\providecommand \bibfield  [0]{\@secondoftwo}%
\providecommand \translation [1]{[#1]}%
\providecommand \BibitemOpen [0]{}%
\providecommand \bibitemStop [0]{}%
\providecommand \bibitemNoStop [0]{.\EOS\space}%
\providecommand \EOS [0]{\spacefactor3000\relax}%
\providecommand \BibitemShut  [1]{\csname bibitem#1\endcsname}%
\let\auto@bib@innerbib\@empty
\bibitem [{\citenamefont {Berthier}\ and\ \citenamefont
  {Biroli}(2011)}]{berthier2011theoretical}%
  \BibitemOpen
  \bibfield  {author} {\bibinfo {author} {\bibfnamefont {L.}~\bibnamefont
  {Berthier}}\ and\ \bibinfo {author} {\bibfnamefont {G.}~\bibnamefont
  {Biroli}},\ }\href
  {https://journals.aps.org/rmp/abstract/10.1103/RevModPhys.83.587} {\bibfield
  {journal} {\bibinfo  {journal} {Reviews of Modern Physics}\ }\textbf
  {\bibinfo {volume} {83}},\ \bibinfo {pages} {587} (\bibinfo {year}
  {2011})}\BibitemShut {NoStop}%
\bibitem [{\citenamefont {Berthier}\ \emph {et~al.}(2011)\citenamefont
  {Berthier}, \citenamefont {Biroli}, \citenamefont {Bouchaud},\ and\
  \citenamefont {Jack}}]{berthier2011overview}%
  \BibitemOpen
  \bibfield  {author} {\bibinfo {author} {\bibfnamefont {L.}~\bibnamefont
  {Berthier}}, \bibinfo {author} {\bibfnamefont {G.}~\bibnamefont {Biroli}},
  \bibinfo {author} {\bibfnamefont {J.-P.}\ \bibnamefont {Bouchaud}}, \ and\
  \bibinfo {author} {\bibfnamefont {R.~L.}\ \bibnamefont {Jack}},\ }\enquote
  {\bibinfo {title} {Overview of different characterizations of dynamic
  heterogeneity},}\ in\ \href@noop {} {\emph {\bibinfo {booktitle} {Dynamical
  Heterogeneities in Glasses, Colloids, and Granular Media}}}\ (\bibinfo
  {publisher} {Oxford University Press},\ \bibinfo {year} {2011})\ pp.\
  \bibinfo {pages} {68--109}\BibitemShut {NoStop}%
\bibitem [{\citenamefont {Berthier}(2011)}]{berthier2011}%
  \BibitemOpen
  \bibfield  {author} {\bibinfo {author} {\bibfnamefont {L.}~\bibnamefont
  {Berthier}},\ }\href {https://physics.aps.org/articles/v4/42} {\bibfield
  {journal} {\bibinfo  {journal} {Physics}\ }\textbf {\bibinfo {volume} {4}},\
  \bibinfo {pages} {42} (\bibinfo {year} {2011})}\BibitemShut {NoStop}%
\bibitem [{\citenamefont {Berthier}\ \emph {et~al.}(2005)\citenamefont
  {Berthier}, \citenamefont {Biroli}, \citenamefont {Bouchaud}, \citenamefont
  {Cipelletti}, \citenamefont {El~Masri}, \citenamefont {L'H{\^o}te},
  \citenamefont {Ladieu},\ and\ \citenamefont {Pierno}}]{berthier2005direct}%
  \BibitemOpen
  \bibfield  {author} {\bibinfo {author} {\bibfnamefont {L.}~\bibnamefont
  {Berthier}}, \bibinfo {author} {\bibfnamefont {G.}~\bibnamefont {Biroli}},
  \bibinfo {author} {\bibfnamefont {J.-P.}\ \bibnamefont {Bouchaud}}, \bibinfo
  {author} {\bibfnamefont {L.}~\bibnamefont {Cipelletti}}, \bibinfo {author}
  {\bibfnamefont {D.}~\bibnamefont {El~Masri}}, \bibinfo {author}
  {\bibfnamefont {D.}~\bibnamefont {L'H{\^o}te}}, \bibinfo {author}
  {\bibfnamefont {F.}~\bibnamefont {Ladieu}}, \ and\ \bibinfo {author}
  {\bibfnamefont {M.}~\bibnamefont {Pierno}},\ }\href
  {https://www.science.org/doi/abs/10.1126/science.1120714} {\bibfield
  {journal} {\bibinfo  {journal} {Science}\ }\textbf {\bibinfo {volume}
  {310}},\ \bibinfo {pages} {1797} (\bibinfo {year} {2005})}\BibitemShut
  {NoStop}%
\bibitem [{\citenamefont {Dalle-Ferrier}\ \emph {et~al.}(2007)\citenamefont
  {Dalle-Ferrier}, \citenamefont {Thibierge}, \citenamefont {Alba-Simionesco},
  \citenamefont {Berthier}, \citenamefont {Biroli}, \citenamefont {Bouchaud},
  \citenamefont {Ladieu}, \citenamefont {L’H{\^o}te},\ and\ \citenamefont
  {Tarjus}}]{dalle2007spatial}%
  \BibitemOpen
  \bibfield  {author} {\bibinfo {author} {\bibfnamefont {C.}~\bibnamefont
  {Dalle-Ferrier}}, \bibinfo {author} {\bibfnamefont {C.}~\bibnamefont
  {Thibierge}}, \bibinfo {author} {\bibfnamefont {C.}~\bibnamefont
  {Alba-Simionesco}}, \bibinfo {author} {\bibfnamefont {L.}~\bibnamefont
  {Berthier}}, \bibinfo {author} {\bibfnamefont {G.}~\bibnamefont {Biroli}},
  \bibinfo {author} {\bibfnamefont {J.-P.}\ \bibnamefont {Bouchaud}}, \bibinfo
  {author} {\bibfnamefont {F.}~\bibnamefont {Ladieu}}, \bibinfo {author}
  {\bibfnamefont {D.}~\bibnamefont {L’H{\^o}te}}, \ and\ \bibinfo {author}
  {\bibfnamefont {G.}~\bibnamefont {Tarjus}},\ }\href
  {https://journals.aps.org/pre/abstract/10.1103/PhysRevE.76.041510} {\bibfield
   {journal} {\bibinfo  {journal} {Physical Review E}\ }\textbf {\bibinfo
  {volume} {76}},\ \bibinfo {pages} {041510} (\bibinfo {year}
  {2007})}\BibitemShut {NoStop}%
\bibitem [{\citenamefont {Kirkpatrick}\ \emph {et~al.}(1989)\citenamefont
  {Kirkpatrick}, \citenamefont {Thirumalai},\ and\ \citenamefont
  {Wolynes}}]{kirkpatrick1989scaling}%
  \BibitemOpen
  \bibfield  {author} {\bibinfo {author} {\bibfnamefont {T.~R.}\ \bibnamefont
  {Kirkpatrick}}, \bibinfo {author} {\bibfnamefont {D.}~\bibnamefont
  {Thirumalai}}, \ and\ \bibinfo {author} {\bibfnamefont {P.~G.}\ \bibnamefont
  {Wolynes}},\ }\href
  {https://journals.aps.org/pra/abstract/10.1103/PhysRevA.40.1045} {\bibfield
  {journal} {\bibinfo  {journal} {Physical Review A}\ }\textbf {\bibinfo
  {volume} {40}},\ \bibinfo {pages} {1045} (\bibinfo {year}
  {1989})}\BibitemShut {NoStop}%
\bibitem [{\citenamefont {Lubchenko}\ and\ \citenamefont
  {Wolynes}(2007)}]{lubchenko2007theory}%
  \BibitemOpen
  \bibfield  {author} {\bibinfo {author} {\bibfnamefont {V.}~\bibnamefont
  {Lubchenko}}\ and\ \bibinfo {author} {\bibfnamefont {P.~G.}\ \bibnamefont
  {Wolynes}},\ }\href
  {https://doi.org/10.1146/annurev.physchem.58.032806.104653} {\bibfield
  {journal} {\bibinfo  {journal} {Annual Review of Physical Chemistry}\
  }\textbf {\bibinfo {volume} {58}},\ \bibinfo {pages} {235} (\bibinfo {year}
  {2007})}\BibitemShut {NoStop}%
\bibitem [{\citenamefont {Wolynes}\ and\ \citenamefont
  {Lubchenko}(2012)}]{wolynes2012structural}%
  \BibitemOpen
  \bibfield  {author} {\bibinfo {author} {\bibfnamefont {P.~G.}\ \bibnamefont
  {Wolynes}}\ and\ \bibinfo {author} {\bibfnamefont {V.}~\bibnamefont
  {Lubchenko}},\ }\href@noop {} {\emph {\bibinfo {title} {Structural glasses
  and supercooled liquids: Theory, experiment, and applications}}}\ (\bibinfo
  {publisher} {John Wiley \& Sons},\ \bibinfo {year} {2012})\BibitemShut
  {NoStop}%
\bibitem [{\citenamefont {Parisi}\ \emph {et~al.}(2020)\citenamefont {Parisi},
  \citenamefont {Urbani},\ and\ \citenamefont {Zamponi}}]{parisi2020theory}%
  \BibitemOpen
  \bibfield  {author} {\bibinfo {author} {\bibfnamefont {G.}~\bibnamefont
  {Parisi}}, \bibinfo {author} {\bibfnamefont {P.}~\bibnamefont {Urbani}}, \
  and\ \bibinfo {author} {\bibfnamefont {F.}~\bibnamefont {Zamponi}},\
  }\href@noop {} {\emph {\bibinfo {title} {Theory of simple glasses: exact
  solutions in infinite dimensions}}}\ (\bibinfo  {publisher} {Cambridge
  University Press},\ \bibinfo {year} {2020})\BibitemShut {NoStop}%
\bibitem [{\citenamefont {Bouchaud}\ and\ \citenamefont
  {Biroli}(2004)}]{bouchaud2004adam}%
  \BibitemOpen
  \bibfield  {author} {\bibinfo {author} {\bibfnamefont {J.-P.}\ \bibnamefont
  {Bouchaud}}\ and\ \bibinfo {author} {\bibfnamefont {G.}~\bibnamefont
  {Biroli}},\ }\href
  {https://aip.scitation.org/doi/abs/10.1063/1.1796231?casa_token=MS19jadsNiAAAAAA:Oxmcd_3ajwuVWE8mOcLhPnMEOXN27Y7vJ5Yf68lgG471SbWMAUJP9AJEaoCQAyAlNpfZbTFsnQQ}
  {\bibfield  {journal} {\bibinfo  {journal} {The Journal of Chemical Physics}\
  }\textbf {\bibinfo {volume} {121}},\ \bibinfo {pages} {7347} (\bibinfo {year}
  {2004})}\BibitemShut {NoStop}%
\bibitem [{\citenamefont {Biroli}\ and\ \citenamefont
  {Bouchaud}(2012)}]{biroli2012random}%
  \BibitemOpen
  \bibfield  {author} {\bibinfo {author} {\bibfnamefont {G.}~\bibnamefont
  {Biroli}}\ and\ \bibinfo {author} {\bibfnamefont {J.-P.}\ \bibnamefont
  {Bouchaud}},\ }\enquote {\bibinfo {title} {The random first-order transition
  theory of glasses: a critical assessment},}\ in\ \href@noop {} {\emph
  {\bibinfo {booktitle} {Structural Glasses and Supercooled Liquids: Theory,
  Experiment, and Applications}}}\ (\bibinfo  {publisher} {John Wiley \&
  Sons},\ \bibinfo {year} {2012})\ pp.\ \bibinfo {pages} {31--113}\BibitemShut
  {NoStop}%
\bibitem [{\citenamefont {Montanari}\ and\ \citenamefont
  {Semerjian}(2006)}]{montanari2006rigorous}%
  \BibitemOpen
  \bibfield  {author} {\bibinfo {author} {\bibfnamefont {A.}~\bibnamefont
  {Montanari}}\ and\ \bibinfo {author} {\bibfnamefont {G.}~\bibnamefont
  {Semerjian}},\ }\href
  {https://link.springer.com/article/10.1007/s10955-006-9175-y} {\bibfield
  {journal} {\bibinfo  {journal} {Journal of Statistical Physics}\ }\textbf
  {\bibinfo {volume} {125}},\ \bibinfo {pages} {23} (\bibinfo {year}
  {2006})}\BibitemShut {NoStop}%
\bibitem [{\citenamefont {Franz}\ and\ \citenamefont
  {Semerjian}(2011)}]{franz2011analytical}%
  \BibitemOpen
  \bibfield  {author} {\bibinfo {author} {\bibfnamefont {S.}~\bibnamefont
  {Franz}}\ and\ \bibinfo {author} {\bibfnamefont {G.}~\bibnamefont
  {Semerjian}},\ }\enquote {\bibinfo {title} {Analytical approaches to time-and
  length scales in models of glasses},}\ in\ \href@noop {} {\emph {\bibinfo
  {booktitle} {Dynamical Heterogeneities in Glasses, Colloids, and Granular
  Media}}}\ (\bibinfo  {publisher} {Oxford University Press},\ \bibinfo {year}
  {2011})\ pp.\ \bibinfo {pages} {407--450}\BibitemShut {NoStop}%
\bibitem [{\citenamefont {Xia}\ and\ \citenamefont
  {Wolynes}(2000)}]{xia2000fragilities}%
  \BibitemOpen
  \bibfield  {author} {\bibinfo {author} {\bibfnamefont {X.}~\bibnamefont
  {Xia}}\ and\ \bibinfo {author} {\bibfnamefont {P.~G.}\ \bibnamefont
  {Wolynes}},\ }\href {https://www.pnas.org/content/97/7/2990.short} {\bibfield
   {journal} {\bibinfo  {journal} {Proceedings of the National Academy of
  Sciences}\ }\textbf {\bibinfo {volume} {97}},\ \bibinfo {pages} {2990}
  (\bibinfo {year} {2000})}\BibitemShut {NoStop}%
\bibitem [{\citenamefont {Castellani}\ and\ \citenamefont
  {Cavagna}(2005)}]{castellani2005spin}%
  \BibitemOpen
  \bibfield  {author} {\bibinfo {author} {\bibfnamefont {T.}~\bibnamefont
  {Castellani}}\ and\ \bibinfo {author} {\bibfnamefont {A.}~\bibnamefont
  {Cavagna}},\ }\href
  {https://iopscience.iop.org/article/10.1088/1742-5468/2005/05/P05012/meta?casa_token=tqUNObEKW4UAAAAA:AwpzcJ0DsAdzuFm_rp-S0Xi3qE14OCHCzz0o_vIrZbY7WtzwvVdPvMrSR8HW87Mf_l11y5zUwAgQ}
  {\bibfield  {journal} {\bibinfo  {journal} {Journal of Statistical Mechanics:
  Theory and Experiment}\ }\textbf {\bibinfo {volume} {2005}},\ \bibinfo
  {pages} {P05012} (\bibinfo {year} {2005})}\BibitemShut {NoStop}%
\bibitem [{\citenamefont {Franz}\ and\ \citenamefont
  {Parisi}(1997)}]{franz1997phase}%
  \BibitemOpen
  \bibfield  {author} {\bibinfo {author} {\bibfnamefont {S.}~\bibnamefont
  {Franz}}\ and\ \bibinfo {author} {\bibfnamefont {G.}~\bibnamefont {Parisi}},\
  }\href {https://journals.aps.org/prl/abstract/10.1103/PhysRevLett.79.2486}
  {\bibfield  {journal} {\bibinfo  {journal} {Physical Review Letters}\
  }\textbf {\bibinfo {volume} {79}},\ \bibinfo {pages} {2486} (\bibinfo {year}
  {1997})}\BibitemShut {NoStop}%
\bibitem [{\citenamefont {Cammarota}\ and\ \citenamefont
  {Biroli}(2012{\natexlab{a}})}]{cammarota2012ideal}%
  \BibitemOpen
  \bibfield  {author} {\bibinfo {author} {\bibfnamefont {C.}~\bibnamefont
  {Cammarota}}\ and\ \bibinfo {author} {\bibfnamefont {G.}~\bibnamefont
  {Biroli}},\ }\href {https://www.pnas.org/content/109/23/8850/} {\bibfield
  {journal} {\bibinfo  {journal} {Proceedings of the National Academy of
  Sciences}\ }\textbf {\bibinfo {volume} {109}},\ \bibinfo {pages} {8850}
  (\bibinfo {year} {2012}{\natexlab{a}})}\BibitemShut {NoStop}%
\bibitem [{\citenamefont {Kob}\ and\ \citenamefont
  {Berthier}(2013)}]{kob2013probing}%
  \BibitemOpen
  \bibfield  {author} {\bibinfo {author} {\bibfnamefont {W.}~\bibnamefont
  {Kob}}\ and\ \bibinfo {author} {\bibfnamefont {L.}~\bibnamefont {Berthier}},\
  }\href {https://journals.aps.org/prl/abstract/10.1103/PhysRevLett.110.245702}
  {\bibfield  {journal} {\bibinfo  {journal} {Physical Review Letters}\
  }\textbf {\bibinfo {volume} {110}},\ \bibinfo {pages} {245702} (\bibinfo
  {year} {2013})}\BibitemShut {NoStop}%
\bibitem [{\citenamefont {Cammarota}\ and\ \citenamefont
  {Biroli}(2013)}]{cammarota2013random}%
  \BibitemOpen
  \bibfield  {author} {\bibinfo {author} {\bibfnamefont {C.}~\bibnamefont
  {Cammarota}}\ and\ \bibinfo {author} {\bibfnamefont {G.}~\bibnamefont
  {Biroli}},\ }\href
  {https://aip.scitation.org/doi/full/10.1063/1.4790400?casa_token=zY_u7k8G4IQAAAAA%3AkyEQBZzTw0wmLCrb4Ar5rd7uaOqsqXh9Pw6LHkn2A-i3kgEGHd0Q-9SaeEb_JcpCmot4-CwgDhI}
  {\bibfield  {journal} {\bibinfo  {journal} {The Journal of Chemical Physics}\
  }\textbf {\bibinfo {volume} {138}},\ \bibinfo {pages} {12A547} (\bibinfo
  {year} {2013})}\BibitemShut {NoStop}%
\bibitem [{\citenamefont {Franz}\ and\ \citenamefont
  {Parisi}(2013)}]{franz2013universality}%
  \BibitemOpen
  \bibfield  {author} {\bibinfo {author} {\bibfnamefont {S.}~\bibnamefont
  {Franz}}\ and\ \bibinfo {author} {\bibfnamefont {G.}~\bibnamefont {Parisi}},\
  }\href
  {https://iopscience.iop.org/article/10.1088/1742-5468/2013/11/P11012/meta}
  {\bibfield  {journal} {\bibinfo  {journal} {Journal of Statistical Mechanics:
  Theory and Experiment}\ }\textbf {\bibinfo {volume} {2013}},\ \bibinfo
  {pages} {P11012} (\bibinfo {year} {2013})}\BibitemShut {NoStop}%
\bibitem [{\citenamefont {Biroli}\ \emph {et~al.}(2014)\citenamefont {Biroli},
  \citenamefont {Cammarota}, \citenamefont {Tarjus},\ and\ \citenamefont
  {Tarzia}}]{biroli2014random}%
  \BibitemOpen
  \bibfield  {author} {\bibinfo {author} {\bibfnamefont {G.}~\bibnamefont
  {Biroli}}, \bibinfo {author} {\bibfnamefont {C.}~\bibnamefont {Cammarota}},
  \bibinfo {author} {\bibfnamefont {G.}~\bibnamefont {Tarjus}}, \ and\ \bibinfo
  {author} {\bibfnamefont {M.}~\bibnamefont {Tarzia}},\ }\href
  {https://journals.aps.org/prl/abstract/10.1103/PhysRevLett.112.175701}
  {\bibfield  {journal} {\bibinfo  {journal} {Physical Review Letters}\
  }\textbf {\bibinfo {volume} {112}},\ \bibinfo {pages} {175701} (\bibinfo
  {year} {2014})}\BibitemShut {NoStop}%
\bibitem [{\citenamefont {Tarjus}(2011)}]{tarjus2011overview}%
  \BibitemOpen
  \bibfield  {author} {\bibinfo {author} {\bibfnamefont {G.}~\bibnamefont
  {Tarjus}},\ }\enquote {\bibinfo {title} {Overview of different
  characterizations of dynamic heterogeneity},}\ in\ \href@noop {} {\emph
  {\bibinfo {booktitle} {Dynamical Heterogeneities in Glasses, Colloids, and
  Granular Media}}}\ (\bibinfo  {publisher} {Oxford University Press},\
  \bibinfo {year} {2011})\ pp.\ \bibinfo {pages} {39--67}\BibitemShut {NoStop}%
\bibitem [{\citenamefont {Ozawa}\ \emph {et~al.}(2019)\citenamefont {Ozawa},
  \citenamefont {Scalliet}, \citenamefont {Ninarello},\ and\ \citenamefont
  {Berthier}}]{ozawa2019does}%
  \BibitemOpen
  \bibfield  {author} {\bibinfo {author} {\bibfnamefont {M.}~\bibnamefont
  {Ozawa}}, \bibinfo {author} {\bibfnamefont {C.}~\bibnamefont {Scalliet}},
  \bibinfo {author} {\bibfnamefont {A.}~\bibnamefont {Ninarello}}, \ and\
  \bibinfo {author} {\bibfnamefont {L.}~\bibnamefont {Berthier}},\ }\href
  {https://aip.scitation.org/doi/full/10.1063/1.5113477?casa_token=tmxN-D5Tl7EAAAAA%3AuXKYLk2gfzrSvkqM3LKKVOJTmh5jEENbr4SKSwI2TYTjv8SDA_TxdNUS2grHpXzLa8x01f4QRRA}
  {\bibfield  {journal} {\bibinfo  {journal} {The Journal of Chemical Physics}\
  }\textbf {\bibinfo {volume} {151}},\ \bibinfo {pages} {084504} (\bibinfo
  {year} {2019})}\BibitemShut {NoStop}%
\bibitem [{\citenamefont {Guiselin}\ \emph {et~al.}(2022)\citenamefont
  {Guiselin}, \citenamefont {Berthier},\ and\ \citenamefont
  {Tarjus}}]{guiselin2021statistical}%
  \BibitemOpen
  \bibfield  {author} {\bibinfo {author} {\bibfnamefont {B.}~\bibnamefont
  {Guiselin}}, \bibinfo {author} {\bibfnamefont {L.}~\bibnamefont {Berthier}},
  \ and\ \bibinfo {author} {\bibfnamefont {G.}~\bibnamefont {Tarjus}},\ }\href
  {\doibase 10.21468/SciPostPhys.12.3.091} {\bibfield  {journal} {\bibinfo
  {journal} {SciPost Phys.}\ }\textbf {\bibinfo {volume} {12}},\ \bibinfo
  {pages} {91} (\bibinfo {year} {2022})}\BibitemShut {NoStop}%
\bibitem [{\citenamefont {Parisi}\ and\ \citenamefont
  {Seoane}(2014)}]{parisi2014liquid}%
  \BibitemOpen
  \bibfield  {author} {\bibinfo {author} {\bibfnamefont {G.}~\bibnamefont
  {Parisi}}\ and\ \bibinfo {author} {\bibfnamefont {B.}~\bibnamefont
  {Seoane}},\ }\href
  {https://journals.aps.org/pre/abstract/10.1103/PhysRevE.89.022309} {\bibfield
   {journal} {\bibinfo  {journal} {Physical Review E}\ }\textbf {\bibinfo
  {volume} {89}},\ \bibinfo {pages} {022309} (\bibinfo {year}
  {2014})}\BibitemShut {NoStop}%
\bibitem [{\citenamefont {Cammarota}\ \emph {et~al.}(2010)\citenamefont
  {Cammarota}, \citenamefont {Cavagna}, \citenamefont {Giardina}, \citenamefont
  {Gradenigo}, \citenamefont {Grigera}, \citenamefont {Parisi},\ and\
  \citenamefont {Verrocchio}}]{cammarota2010phase}%
  \BibitemOpen
  \bibfield  {author} {\bibinfo {author} {\bibfnamefont {C.}~\bibnamefont
  {Cammarota}}, \bibinfo {author} {\bibfnamefont {A.}~\bibnamefont {Cavagna}},
  \bibinfo {author} {\bibfnamefont {I.}~\bibnamefont {Giardina}}, \bibinfo
  {author} {\bibfnamefont {G.}~\bibnamefont {Gradenigo}}, \bibinfo {author}
  {\bibfnamefont {T.~S.}\ \bibnamefont {Grigera}}, \bibinfo {author}
  {\bibfnamefont {G.}~\bibnamefont {Parisi}}, \ and\ \bibinfo {author}
  {\bibfnamefont {P.}~\bibnamefont {Verrocchio}},\ }\href
  {https://journals.aps.org/prl/abstract/10.1103/PhysRevLett.105.055703}
  {\bibfield  {journal} {\bibinfo  {journal} {Physical Review Letters}\
  }\textbf {\bibinfo {volume} {105}},\ \bibinfo {pages} {055703} (\bibinfo
  {year} {2010})}\BibitemShut {NoStop}%
\bibitem [{\citenamefont {Berthier}(2013)}]{berthier2013overlap}%
  \BibitemOpen
  \bibfield  {author} {\bibinfo {author} {\bibfnamefont {L.}~\bibnamefont
  {Berthier}},\ }\href
  {https://journals.aps.org/pre/abstract/10.1103/PhysRevE.88.022313} {\bibfield
   {journal} {\bibinfo  {journal} {Physical Review E}\ }\textbf {\bibinfo
  {volume} {88}},\ \bibinfo {pages} {022313} (\bibinfo {year}
  {2013})}\BibitemShut {NoStop}%
\bibitem [{\citenamefont {Berthier}\ and\ \citenamefont
  {Coslovich}(2014)}]{berthier2014novel}%
  \BibitemOpen
  \bibfield  {author} {\bibinfo {author} {\bibfnamefont {L.}~\bibnamefont
  {Berthier}}\ and\ \bibinfo {author} {\bibfnamefont {D.}~\bibnamefont
  {Coslovich}},\ }\href {https://www.pnas.org/content/111/32/11668.short}
  {\bibfield  {journal} {\bibinfo  {journal} {Proceedings of the National
  Academy of Sciences}\ }\textbf {\bibinfo {volume} {111}},\ \bibinfo {pages}
  {11668} (\bibinfo {year} {2014})}\BibitemShut {NoStop}%
\bibitem [{\citenamefont {Cammarota}\ and\ \citenamefont
  {Seoane}(2016)}]{cammarota2016first}%
  \BibitemOpen
  \bibfield  {author} {\bibinfo {author} {\bibfnamefont {C.}~\bibnamefont
  {Cammarota}}\ and\ \bibinfo {author} {\bibfnamefont {B.}~\bibnamefont
  {Seoane}},\ }\href
  {https://journals.aps.org/prb/abstract/10.1103/PhysRevB.94.180201} {\bibfield
   {journal} {\bibinfo  {journal} {Physical Review B}\ }\textbf {\bibinfo
  {volume} {94}},\ \bibinfo {pages} {180201} (\bibinfo {year}
  {2016})}\BibitemShut {NoStop}%
\bibitem [{\citenamefont {Berthier}\ and\ \citenamefont
  {Jack}(2015)}]{berthier2015evidence}%
  \BibitemOpen
  \bibfield  {author} {\bibinfo {author} {\bibfnamefont {L.}~\bibnamefont
  {Berthier}}\ and\ \bibinfo {author} {\bibfnamefont {R.~L.}\ \bibnamefont
  {Jack}},\ }\href
  {https://journals.aps.org/prl/abstract/10.1103/PhysRevLett.114.205701}
  {\bibfield  {journal} {\bibinfo  {journal} {Physical Review Letters}\
  }\textbf {\bibinfo {volume} {114}},\ \bibinfo {pages} {205701} (\bibinfo
  {year} {2015})}\BibitemShut {NoStop}%
\bibitem [{\citenamefont {Jack}\ and\ \citenamefont
  {Garrahan}(2016)}]{jack2016phase}%
  \BibitemOpen
  \bibfield  {author} {\bibinfo {author} {\bibfnamefont {R.~L.}\ \bibnamefont
  {Jack}}\ and\ \bibinfo {author} {\bibfnamefont {J.~P.}\ \bibnamefont
  {Garrahan}},\ }\href
  {https://journals.aps.org/prl/abstract/10.1103/PhysRevLett.116.055702}
  {\bibfield  {journal} {\bibinfo  {journal} {Physical Review Letters}\
  }\textbf {\bibinfo {volume} {116}},\ \bibinfo {pages} {055702} (\bibinfo
  {year} {2016})}\BibitemShut {NoStop}%
\bibitem [{\citenamefont {Cavagna}\ \emph {et~al.}(2007)\citenamefont
  {Cavagna}, \citenamefont {Grigera},\ and\ \citenamefont
  {Verrocchio}}]{cavagna2007mosaic}%
  \BibitemOpen
  \bibfield  {author} {\bibinfo {author} {\bibfnamefont {A.}~\bibnamefont
  {Cavagna}}, \bibinfo {author} {\bibfnamefont {T.~S.}\ \bibnamefont
  {Grigera}}, \ and\ \bibinfo {author} {\bibfnamefont {P.}~\bibnamefont
  {Verrocchio}},\ }\href
  {https://journals.aps.org/prl/abstract/10.1103/PhysRevLett.98.187801}
  {\bibfield  {journal} {\bibinfo  {journal} {Physical Review Letters}\
  }\textbf {\bibinfo {volume} {98}},\ \bibinfo {pages} {187801} (\bibinfo
  {year} {2007})}\BibitemShut {NoStop}%
\bibitem [{\citenamefont {Biroli}\ \emph {et~al.}(2008)\citenamefont {Biroli},
  \citenamefont {Bouchaud}, \citenamefont {Cavagna}, \citenamefont {Grigera},\
  and\ \citenamefont {Verrocchio}}]{biroli2008thermodynamic}%
  \BibitemOpen
  \bibfield  {author} {\bibinfo {author} {\bibfnamefont {G.}~\bibnamefont
  {Biroli}}, \bibinfo {author} {\bibfnamefont {J.-P.}\ \bibnamefont
  {Bouchaud}}, \bibinfo {author} {\bibfnamefont {A.}~\bibnamefont {Cavagna}},
  \bibinfo {author} {\bibfnamefont {T.~S.}\ \bibnamefont {Grigera}}, \ and\
  \bibinfo {author} {\bibfnamefont {P.}~\bibnamefont {Verrocchio}},\ }\href
  {https://www.nature.com/articles/nphys1050} {\bibfield  {journal} {\bibinfo
  {journal} {Nature Physics}\ }\textbf {\bibinfo {volume} {4}},\ \bibinfo
  {pages} {771} (\bibinfo {year} {2008})}\BibitemShut {NoStop}%
\bibitem [{\citenamefont {Nagamanasa}\ \emph {et~al.}(2015)\citenamefont
  {Nagamanasa}, \citenamefont {Gokhale}, \citenamefont {Sood},\ and\
  \citenamefont {Ganapathy}}]{nagamanasa2015direct}%
  \BibitemOpen
  \bibfield  {author} {\bibinfo {author} {\bibfnamefont {K.~H.}\ \bibnamefont
  {Nagamanasa}}, \bibinfo {author} {\bibfnamefont {S.}~\bibnamefont {Gokhale}},
  \bibinfo {author} {\bibfnamefont {A.}~\bibnamefont {Sood}}, \ and\ \bibinfo
  {author} {\bibfnamefont {R.}~\bibnamefont {Ganapathy}},\ }\href
  {https://www.nature.com/articles/nphys3289} {\bibfield  {journal} {\bibinfo
  {journal} {Nature Physics}\ }\textbf {\bibinfo {volume} {11}},\ \bibinfo
  {pages} {403} (\bibinfo {year} {2015})}\BibitemShut {NoStop}%
\bibitem [{\citenamefont {Yaida}\ \emph {et~al.}(2016)\citenamefont {Yaida},
  \citenamefont {Berthier}, \citenamefont {Charbonneau},\ and\ \citenamefont
  {Tarjus}}]{yaida2016point}%
  \BibitemOpen
  \bibfield  {author} {\bibinfo {author} {\bibfnamefont {S.}~\bibnamefont
  {Yaida}}, \bibinfo {author} {\bibfnamefont {L.}~\bibnamefont {Berthier}},
  \bibinfo {author} {\bibfnamefont {P.}~\bibnamefont {Charbonneau}}, \ and\
  \bibinfo {author} {\bibfnamefont {G.}~\bibnamefont {Tarjus}},\ }\href
  {https://journals.aps.org/pre/abstract/10.1103/PhysRevE.94.032605} {\bibfield
   {journal} {\bibinfo  {journal} {Physical Review E}\ }\textbf {\bibinfo
  {volume} {94}},\ \bibinfo {pages} {032605} (\bibinfo {year}
  {2016})}\BibitemShut {NoStop}%
\bibitem [{\citenamefont {Berthier}\ \emph {et~al.}(2016)\citenamefont
  {Berthier}, \citenamefont {Charbonneau},\ and\ \citenamefont
  {Yaida}}]{berthier2016efficient}%
  \BibitemOpen
  \bibfield  {author} {\bibinfo {author} {\bibfnamefont {L.}~\bibnamefont
  {Berthier}}, \bibinfo {author} {\bibfnamefont {P.}~\bibnamefont
  {Charbonneau}}, \ and\ \bibinfo {author} {\bibfnamefont {S.}~\bibnamefont
  {Yaida}},\ }\href
  {https://aip.scitation.org/doi/full/10.1063/1.4939640?casa_token=t0QANSCCpCcAAAAA%3ASmmU_A3OTTJx12Gmu6zfMazpuaZgzVhVlVhEcDDs-UIge1X0i1gnNbt26PzvfOl5hW_vCbuWv7k}
  {\bibfield  {journal} {\bibinfo  {journal} {The Journal of Chemical Physics}\
  }\textbf {\bibinfo {volume} {144}},\ \bibinfo {pages} {024501} (\bibinfo
  {year} {2016})}\BibitemShut {NoStop}%
\bibitem [{\citenamefont {Berthier}\ \emph
  {et~al.}(2019{\natexlab{a}})\citenamefont {Berthier}, \citenamefont
  {Charbonneau}, \citenamefont {Ninarello}, \citenamefont {Ozawa},\ and\
  \citenamefont {Yaida}}]{berthier2019zero}%
  \BibitemOpen
  \bibfield  {author} {\bibinfo {author} {\bibfnamefont {L.}~\bibnamefont
  {Berthier}}, \bibinfo {author} {\bibfnamefont {P.}~\bibnamefont
  {Charbonneau}}, \bibinfo {author} {\bibfnamefont {A.}~\bibnamefont
  {Ninarello}}, \bibinfo {author} {\bibfnamefont {M.}~\bibnamefont {Ozawa}}, \
  and\ \bibinfo {author} {\bibfnamefont {S.}~\bibnamefont {Yaida}},\ }\href
  {https://www.nature.com/articles/s41467-019-09512-3} {\bibfield  {journal}
  {\bibinfo  {journal} {Nature Communications}\ }\textbf {\bibinfo {volume}
  {10}},\ \bibinfo {pages} {1} (\bibinfo {year}
  {2019}{\natexlab{a}})}\BibitemShut {NoStop}%
\bibitem [{\citenamefont {Guiselin}\ \emph
  {et~al.}(2020{\natexlab{a}})\citenamefont {Guiselin}, \citenamefont
  {Berthier},\ and\ \citenamefont {Tarjus}}]{guiselin2020random}%
  \BibitemOpen
  \bibfield  {author} {\bibinfo {author} {\bibfnamefont {B.}~\bibnamefont
  {Guiselin}}, \bibinfo {author} {\bibfnamefont {L.}~\bibnamefont {Berthier}},
  \ and\ \bibinfo {author} {\bibfnamefont {G.}~\bibnamefont {Tarjus}},\ }\href
  {https://journals.aps.org/pre/abstract/10.1103/PhysRevE.102.042129}
  {\bibfield  {journal} {\bibinfo  {journal} {Physical Review E}\ }\textbf
  {\bibinfo {volume} {102}},\ \bibinfo {pages} {042129} (\bibinfo {year}
  {2020}{\natexlab{a}})}\BibitemShut {NoStop}%
\bibitem [{\citenamefont {Cammarota}\ \emph
  {et~al.}(2009{\natexlab{a}})\citenamefont {Cammarota}, \citenamefont
  {Cavagna}, \citenamefont {Gradenigo}, \citenamefont {Grigera},\ and\
  \citenamefont {Verrocchio}}]{cammarota2009evidence}%
  \BibitemOpen
  \bibfield  {author} {\bibinfo {author} {\bibfnamefont {C.}~\bibnamefont
  {Cammarota}}, \bibinfo {author} {\bibfnamefont {A.}~\bibnamefont {Cavagna}},
  \bibinfo {author} {\bibfnamefont {G.}~\bibnamefont {Gradenigo}}, \bibinfo
  {author} {\bibfnamefont {T.}~\bibnamefont {Grigera}}, \ and\ \bibinfo
  {author} {\bibfnamefont {P.}~\bibnamefont {Verrocchio}},\ }\href
  {https://iopscience.iop.org/article/10.1088/1742-5468/2009/12/L12002/meta?casa_token=Jm9c6eLP_3sAAAAA:1rr95BCw5S18I7JfjRLAnTNhBMK7a_BCRzx2ptbJXG9haE6vuuqtjxNjRDa2ZQ2ijD39UkVU8tfd}
  {\bibfield  {journal} {\bibinfo  {journal} {Journal of Statistical Mechanics:
  Theory and Experiment}\ }\textbf {\bibinfo {volume} {2009}},\ \bibinfo
  {pages} {L12002} (\bibinfo {year} {2009}{\natexlab{a}})}\BibitemShut
  {NoStop}%
\bibitem [{\citenamefont {Cammarota}\ \emph
  {et~al.}(2009{\natexlab{b}})\citenamefont {Cammarota}, \citenamefont
  {Cavagna}, \citenamefont {Gradenigo}, \citenamefont {Grigera},\ and\
  \citenamefont {Verrocchio}}]{cammarota2009numerical}%
  \BibitemOpen
  \bibfield  {author} {\bibinfo {author} {\bibfnamefont {C.}~\bibnamefont
  {Cammarota}}, \bibinfo {author} {\bibfnamefont {A.}~\bibnamefont {Cavagna}},
  \bibinfo {author} {\bibfnamefont {G.}~\bibnamefont {Gradenigo}}, \bibinfo
  {author} {\bibfnamefont {T.~S.}\ \bibnamefont {Grigera}}, \ and\ \bibinfo
  {author} {\bibfnamefont {P.}~\bibnamefont {Verrocchio}},\ }\href
  {https://aip.scitation.org/doi/full/10.1063/1.3257739?casa_token=_t3RmmFDXzUAAAAA%3A-cbO_XXESd9l9ziOSUu1pXlDg-bWlfhMHJkq77YAHLXLVYct-wUTtog4CVS1MxHE7aPev68D4zY}
  {\bibfield  {journal} {\bibinfo  {journal} {The Journal of Chemical Physics}\
  }\textbf {\bibinfo {volume} {131}},\ \bibinfo {pages} {194901} (\bibinfo
  {year} {2009}{\natexlab{b}})}\BibitemShut {NoStop}%
\bibitem [{\citenamefont {Ganapathi}\ \emph {et~al.}(2018)\citenamefont
  {Ganapathi}, \citenamefont {Nagamanasa}, \citenamefont {Sood},\ and\
  \citenamefont {Ganapathy}}]{ganapathi2018measurements}%
  \BibitemOpen
  \bibfield  {author} {\bibinfo {author} {\bibfnamefont {D.}~\bibnamefont
  {Ganapathi}}, \bibinfo {author} {\bibfnamefont {K.~H.}\ \bibnamefont
  {Nagamanasa}}, \bibinfo {author} {\bibfnamefont {A.}~\bibnamefont {Sood}}, \
  and\ \bibinfo {author} {\bibfnamefont {R.}~\bibnamefont {Ganapathy}},\ }\href
  {https://www.nature.com/articles/s41467-018-02836-6} {\bibfield  {journal}
  {\bibinfo  {journal} {Nature Communications}\ }\textbf {\bibinfo {volume}
  {9}},\ \bibinfo {pages} {1} (\bibinfo {year} {2018})}\BibitemShut {NoStop}%
\bibitem [{\citenamefont {Berthier}\ \emph
  {et~al.}(2019{\natexlab{b}})\citenamefont {Berthier}, \citenamefont {Ozawa},\
  and\ \citenamefont {Scalliet}}]{berthier2019configurational}%
  \BibitemOpen
  \bibfield  {author} {\bibinfo {author} {\bibfnamefont {L.}~\bibnamefont
  {Berthier}}, \bibinfo {author} {\bibfnamefont {M.}~\bibnamefont {Ozawa}}, \
  and\ \bibinfo {author} {\bibfnamefont {C.}~\bibnamefont {Scalliet}},\ }\href
  {https://aip.scitation.org/doi/full/10.1063/1.5091961} {\bibfield  {journal}
  {\bibinfo  {journal} {The Journal of Chemical Physics}\ }\textbf {\bibinfo
  {volume} {150}},\ \bibinfo {pages} {160902} (\bibinfo {year}
  {2019}{\natexlab{b}})}\BibitemShut {NoStop}%
\bibitem [{\citenamefont {Sastry}(2001)}]{sastry2001relationship}%
  \BibitemOpen
  \bibfield  {author} {\bibinfo {author} {\bibfnamefont {S.}~\bibnamefont
  {Sastry}},\ }\href {https://www.nature.com/articles/35051524} {\bibfield
  {journal} {\bibinfo  {journal} {Nature}\ }\textbf {\bibinfo {volume} {409}},\
  \bibinfo {pages} {164} (\bibinfo {year} {2001})}\BibitemShut {NoStop}%
\bibitem [{\citenamefont {Sengupta}\ \emph {et~al.}(2012)\citenamefont
  {Sengupta}, \citenamefont {Karmakar}, \citenamefont {Dasgupta},\ and\
  \citenamefont {Sastry}}]{sengupta2012adam}%
  \BibitemOpen
  \bibfield  {author} {\bibinfo {author} {\bibfnamefont {S.}~\bibnamefont
  {Sengupta}}, \bibinfo {author} {\bibfnamefont {S.}~\bibnamefont {Karmakar}},
  \bibinfo {author} {\bibfnamefont {C.}~\bibnamefont {Dasgupta}}, \ and\
  \bibinfo {author} {\bibfnamefont {S.}~\bibnamefont {Sastry}},\ }\href
  {https://journals.aps.org/prl/abstract/10.1103/PhysRevLett.109.095705}
  {\bibfield  {journal} {\bibinfo  {journal} {Physical Review Letters}\
  }\textbf {\bibinfo {volume} {109}},\ \bibinfo {pages} {095705} (\bibinfo
  {year} {2012})}\BibitemShut {NoStop}%
\bibitem [{\citenamefont {Berthier}\ \emph {et~al.}(2017)\citenamefont
  {Berthier}, \citenamefont {Charbonneau}, \citenamefont {Coslovich},
  \citenamefont {Ninarello}, \citenamefont {Ozawa},\ and\ \citenamefont
  {Yaida}}]{berthier2017configurational}%
  \BibitemOpen
  \bibfield  {author} {\bibinfo {author} {\bibfnamefont {L.}~\bibnamefont
  {Berthier}}, \bibinfo {author} {\bibfnamefont {P.}~\bibnamefont
  {Charbonneau}}, \bibinfo {author} {\bibfnamefont {D.}~\bibnamefont
  {Coslovich}}, \bibinfo {author} {\bibfnamefont {A.}~\bibnamefont
  {Ninarello}}, \bibinfo {author} {\bibfnamefont {M.}~\bibnamefont {Ozawa}}, \
  and\ \bibinfo {author} {\bibfnamefont {S.}~\bibnamefont {Yaida}},\ }\href
  {https://www.pnas.org/content/114/43/11356.short} {\bibfield  {journal}
  {\bibinfo  {journal} {Proceedings of the National Academy of Sciences}\
  }\textbf {\bibinfo {volume} {114}},\ \bibinfo {pages} {11356} (\bibinfo
  {year} {2017})}\BibitemShut {NoStop}%
\bibitem [{\citenamefont {Ozawa}\ \emph {et~al.}(2018)\citenamefont {Ozawa},
  \citenamefont {Parisi},\ and\ \citenamefont
  {Berthier}}]{ozawa2018configurational}%
  \BibitemOpen
  \bibfield  {author} {\bibinfo {author} {\bibfnamefont {M.}~\bibnamefont
  {Ozawa}}, \bibinfo {author} {\bibfnamefont {G.}~\bibnamefont {Parisi}}, \
  and\ \bibinfo {author} {\bibfnamefont {L.}~\bibnamefont {Berthier}},\ }\href
  {https://aip.scitation.org/doi/full/10.1063/1.5040975?casa_token=_RCfsrw2JjcAAAAA%3Aa5iLqz3Nk7T22PeN3nQYsiwCbgFvd7vlbIoo7cAQ6uGuzHLL2HkfNlME-eIaW1ryvg0JuRnMzJY}
  {\bibfield  {journal} {\bibinfo  {journal} {The Journal of Chemical Physics}\
  }\textbf {\bibinfo {volume} {149}},\ \bibinfo {pages} {154501} (\bibinfo
  {year} {2018})}\BibitemShut {NoStop}%
\bibitem [{\citenamefont {Cammarota}\ \emph {et~al.}(2011)\citenamefont
  {Cammarota}, \citenamefont {Biroli}, \citenamefont {Tarzia},\ and\
  \citenamefont {Tarjus}}]{cammarota2011renormalization}%
  \BibitemOpen
  \bibfield  {author} {\bibinfo {author} {\bibfnamefont {C.}~\bibnamefont
  {Cammarota}}, \bibinfo {author} {\bibfnamefont {G.}~\bibnamefont {Biroli}},
  \bibinfo {author} {\bibfnamefont {M.}~\bibnamefont {Tarzia}}, \ and\ \bibinfo
  {author} {\bibfnamefont {G.}~\bibnamefont {Tarjus}},\ }\href
  {https://journals.aps.org/prl/abstract/10.1103/PhysRevLett.106.115705}
  {\bibfield  {journal} {\bibinfo  {journal} {Physical Review Letters}\
  }\textbf {\bibinfo {volume} {106}},\ \bibinfo {pages} {115705} (\bibinfo
  {year} {2011})}\BibitemShut {NoStop}%
\bibitem [{\citenamefont {Cammarota}\ and\ \citenamefont
  {Biroli}(2012{\natexlab{b}})}]{cammarota2012patch}%
  \BibitemOpen
  \bibfield  {author} {\bibinfo {author} {\bibfnamefont {C.}~\bibnamefont
  {Cammarota}}\ and\ \bibinfo {author} {\bibfnamefont {G.}~\bibnamefont
  {Biroli}},\ }\href
  {https://iopscience.iop.org/article/10.1209/0295-5075/98/36005/meta?casa_token=OU71RwwvUX0AAAAA:U5gJGd0bEdoQL7sKiF4AIkWMx8mUtM0e0sb2dI_p6P2luP_lOR0qmrYEXKKM_M7UvSn6v0s9Qjxu}
  {\bibfield  {journal} {\bibinfo  {journal} {EPL (Europhysics Letters)}\
  }\textbf {\bibinfo {volume} {98}},\ \bibinfo {pages} {36005} (\bibinfo {year}
  {2012}{\natexlab{b}})}\BibitemShut {NoStop}%
\bibitem [{\citenamefont {Xia}\ and\ \citenamefont
  {Wolynes}(2001)}]{xia2001microscopic}%
  \BibitemOpen
  \bibfield  {author} {\bibinfo {author} {\bibfnamefont {X.}~\bibnamefont
  {Xia}}\ and\ \bibinfo {author} {\bibfnamefont {P.~G.}\ \bibnamefont
  {Wolynes}},\ }\href
  {https://journals.aps.org/prl/abstract/10.1103/PhysRevLett.86.5526}
  {\bibfield  {journal} {\bibinfo  {journal} {Physical Review Letters}\
  }\textbf {\bibinfo {volume} {86}},\ \bibinfo {pages} {5526} (\bibinfo {year}
  {2001})}\BibitemShut {NoStop}%
\bibitem [{\citenamefont {Lubchenko}\ and\ \citenamefont
  {Wolynes}(2004)}]{lubchenko2004theory}%
  \BibitemOpen
  \bibfield  {author} {\bibinfo {author} {\bibfnamefont {V.}~\bibnamefont
  {Lubchenko}}\ and\ \bibinfo {author} {\bibfnamefont {P.~G.}\ \bibnamefont
  {Wolynes}},\ }\href
  {https://aip.scitation.org/doi/abs/10.1063/1.1771633?casa_token=aGASDYRQmtoAAAAA:7EgyPhzGGvwTpzEEWRKOgwn9KHGu1Yn7rNsWpzxiQE6DKufDqAnWavXzf_X7F_Nz31f-V1jBdrQ}
  {\bibfield  {journal} {\bibinfo  {journal} {The Journal of Chemical Physics}\
  }\textbf {\bibinfo {volume} {121}},\ \bibinfo {pages} {2852} (\bibinfo {year}
  {2004})}\BibitemShut {NoStop}%
\bibitem [{\citenamefont {Dzero}\ \emph {et~al.}(2009)\citenamefont {Dzero},
  \citenamefont {Schmalian},\ and\ \citenamefont {Wolynes}}]{dzero2009replica}%
  \BibitemOpen
  \bibfield  {author} {\bibinfo {author} {\bibfnamefont {M.}~\bibnamefont
  {Dzero}}, \bibinfo {author} {\bibfnamefont {J.}~\bibnamefont {Schmalian}}, \
  and\ \bibinfo {author} {\bibfnamefont {P.~G.}\ \bibnamefont {Wolynes}},\
  }\href {https://journals.aps.org/prb/abstract/10.1103/PhysRevB.80.024204}
  {\bibfield  {journal} {\bibinfo  {journal} {Physical Review B}\ }\textbf
  {\bibinfo {volume} {80}},\ \bibinfo {pages} {024204} (\bibinfo {year}
  {2009})}\BibitemShut {NoStop}%
\bibitem [{\citenamefont {Bouchaud}\ and\ \citenamefont
  {M{\'e}zard}(1994)}]{bouchaud1994self}%
  \BibitemOpen
  \bibfield  {author} {\bibinfo {author} {\bibfnamefont {J.-P.}\ \bibnamefont
  {Bouchaud}}\ and\ \bibinfo {author} {\bibfnamefont {M.}~\bibnamefont
  {M{\'e}zard}},\ }\href
  {https://jp1.journaldephysique.org/articles/jp1/abs/1994/08/jp1v4p1109/jp1v4p1109.html}
  {\bibfield  {journal} {\bibinfo  {journal} {Journal de Physique I}\ }\textbf
  {\bibinfo {volume} {4}},\ \bibinfo {pages} {1109} (\bibinfo {year}
  {1994})}\BibitemShut {NoStop}%
\bibitem [{\citenamefont {Stevenson}\ \emph {et~al.}(2008)\citenamefont
  {Stevenson}, \citenamefont {Walczak}, \citenamefont {Hall},\ and\
  \citenamefont {Wolynes}}]{stevenson2008constructing}%
  \BibitemOpen
  \bibfield  {author} {\bibinfo {author} {\bibfnamefont {J.~D.}\ \bibnamefont
  {Stevenson}}, \bibinfo {author} {\bibfnamefont {A.~M.}\ \bibnamefont
  {Walczak}}, \bibinfo {author} {\bibfnamefont {R.~W.}\ \bibnamefont {Hall}}, \
  and\ \bibinfo {author} {\bibfnamefont {P.~G.}\ \bibnamefont {Wolynes}},\
  }\href
  {https://aip.scitation.org/doi/full/10.1063/1.3009827?casa_token=qP02PT7WtrcAAAAA%3AQFRoNmrdFU4I2-Z3ZTe-uwg5kFr-9yBiVEMOyGR249nhezLwyGWJ-TgTjvoLjBHIWKomsqZruyw}
  {\bibfield  {journal} {\bibinfo  {journal} {The Journal of Chemical Physics}\
  }\textbf {\bibinfo {volume} {129}},\ \bibinfo {pages} {194505} (\bibinfo
  {year} {2008})}\BibitemShut {NoStop}%
\bibitem [{\citenamefont {Biroli}\ \emph
  {et~al.}(2018{\natexlab{a}})\citenamefont {Biroli}, \citenamefont
  {Cammarota}, \citenamefont {Tarjus},\ and\ \citenamefont
  {Tarzia}}]{biroli2018random1}%
  \BibitemOpen
  \bibfield  {author} {\bibinfo {author} {\bibfnamefont {G.}~\bibnamefont
  {Biroli}}, \bibinfo {author} {\bibfnamefont {C.}~\bibnamefont {Cammarota}},
  \bibinfo {author} {\bibfnamefont {G.}~\bibnamefont {Tarjus}}, \ and\ \bibinfo
  {author} {\bibfnamefont {M.}~\bibnamefont {Tarzia}},\ }\href
  {https://journals.aps.org/prb/abstract/10.1103/PhysRevB.98.174205} {\bibfield
   {journal} {\bibinfo  {journal} {Physical Review B}\ }\textbf {\bibinfo
  {volume} {98}},\ \bibinfo {pages} {174205} (\bibinfo {year}
  {2018}{\natexlab{a}})}\BibitemShut {NoStop}%
\bibitem [{\citenamefont {Biroli}\ \emph
  {et~al.}(2018{\natexlab{b}})\citenamefont {Biroli}, \citenamefont
  {Cammarota}, \citenamefont {Tarjus},\ and\ \citenamefont
  {Tarzia}}]{biroli2018random2}%
  \BibitemOpen
  \bibfield  {author} {\bibinfo {author} {\bibfnamefont {G.}~\bibnamefont
  {Biroli}}, \bibinfo {author} {\bibfnamefont {C.}~\bibnamefont {Cammarota}},
  \bibinfo {author} {\bibfnamefont {G.}~\bibnamefont {Tarjus}}, \ and\ \bibinfo
  {author} {\bibfnamefont {M.}~\bibnamefont {Tarzia}},\ }\href
  {https://journals.aps.org/prb/abstract/10.1103/PhysRevB.98.174206} {\bibfield
   {journal} {\bibinfo  {journal} {Physical Review B}\ }\textbf {\bibinfo
  {volume} {98}},\ \bibinfo {pages} {174206} (\bibinfo {year}
  {2018}{\natexlab{b}})}\BibitemShut {NoStop}%
\bibitem [{\citenamefont {Berthier}(2021)}]{PhysRevLett.127.088002}%
  \BibitemOpen
  \bibfield  {author} {\bibinfo {author} {\bibfnamefont {L.}~\bibnamefont
  {Berthier}},\ }\href {\doibase 10.1103/PhysRevLett.127.088002} {\bibfield
  {journal} {\bibinfo  {journal} {Physical Review Letters}\ }\textbf {\bibinfo
  {volume} {127}},\ \bibinfo {pages} {088002} (\bibinfo {year}
  {2021})}\BibitemShut {NoStop}%
\bibitem [{\citenamefont {Mizuno}\ \emph {et~al.}(2013)\citenamefont {Mizuno},
  \citenamefont {Mossa},\ and\ \citenamefont {Barrat}}]{mizuno2013measuring}%
  \BibitemOpen
  \bibfield  {author} {\bibinfo {author} {\bibfnamefont {H.}~\bibnamefont
  {Mizuno}}, \bibinfo {author} {\bibfnamefont {S.}~\bibnamefont {Mossa}}, \
  and\ \bibinfo {author} {\bibfnamefont {J.-L.}\ \bibnamefont {Barrat}},\
  }\href {https://journals.aps.org/pre/abstract/10.1103/PhysRevE.87.042306}
  {\bibfield  {journal} {\bibinfo  {journal} {Physical Review E}\ }\textbf
  {\bibinfo {volume} {87}},\ \bibinfo {pages} {042306} (\bibinfo {year}
  {2013})}\BibitemShut {NoStop}%
\bibitem [{\citenamefont {Shakerpoor}\ \emph {et~al.}(2020)\citenamefont
  {Shakerpoor}, \citenamefont {Flenner},\ and\ \citenamefont
  {Szamel}}]{shakerpoor2020stability}%
  \BibitemOpen
  \bibfield  {author} {\bibinfo {author} {\bibfnamefont {A.}~\bibnamefont
  {Shakerpoor}}, \bibinfo {author} {\bibfnamefont {E.}~\bibnamefont {Flenner}},
  \ and\ \bibinfo {author} {\bibfnamefont {G.}~\bibnamefont {Szamel}},\ }\href
  {https://pubs.rsc.org/en/content/articlehtml/2020/sm/c9sm02022e?casa_token=dXIlvQ3mEk8AAAAA:0JHe7UOpjG4Rr1PHTVUr-Q67LTTBMJCn5WawqN7xazf14cLjKNxZ6OFCHf7KP358ahlNc4AP7Pqrfg}
  {\bibfield  {journal} {\bibinfo  {journal} {Soft Matter}\ }\textbf {\bibinfo
  {volume} {16}},\ \bibinfo {pages} {914} (\bibinfo {year} {2020})}\BibitemShut
  {NoStop}%
\bibitem [{\citenamefont {Widmer-Cooper}\ \emph {et~al.}(2008)\citenamefont
  {Widmer-Cooper}, \citenamefont {Perry}, \citenamefont {Harrowell},\ and\
  \citenamefont {Reichman}}]{widmer2008irreversible}%
  \BibitemOpen
  \bibfield  {author} {\bibinfo {author} {\bibfnamefont {A.}~\bibnamefont
  {Widmer-Cooper}}, \bibinfo {author} {\bibfnamefont {H.}~\bibnamefont
  {Perry}}, \bibinfo {author} {\bibfnamefont {P.}~\bibnamefont {Harrowell}}, \
  and\ \bibinfo {author} {\bibfnamefont {D.~R.}\ \bibnamefont {Reichman}},\
  }\href {https://www.nature.com/articles/nphys1025} {\bibfield  {journal}
  {\bibinfo  {journal} {Nature Physics}\ }\textbf {\bibinfo {volume} {4}},\
  \bibinfo {pages} {711} (\bibinfo {year} {2008})}\BibitemShut {NoStop}%
\bibitem [{\citenamefont {Lerner}\ and\ \citenamefont
  {Bouchbinder}(2018)}]{doi:10.1063/1.5024776}%
  \BibitemOpen
  \bibfield  {author} {\bibinfo {author} {\bibfnamefont {E.}~\bibnamefont
  {Lerner}}\ and\ \bibinfo {author} {\bibfnamefont {E.}~\bibnamefont
  {Bouchbinder}},\ }\href {https://doi.org/10.1063/1.5024776} {\bibfield
  {journal} {\bibinfo  {journal} {The Journal of Chemical Physics}\ }\textbf
  {\bibinfo {volume} {148}},\ \bibinfo {pages} {214502} (\bibinfo {year}
  {2018})}\BibitemShut {NoStop}%
\bibitem [{\citenamefont {Barbot}\ \emph {et~al.}(2018)\citenamefont {Barbot},
  \citenamefont {Lerbinger}, \citenamefont {Hernandez-Garcia}, \citenamefont
  {Garc{\'\i}a-Garc{\'\i}a}, \citenamefont {Falk}, \citenamefont
  {Vandembroucq},\ and\ \citenamefont {Patinet}}]{barbot2018local}%
  \BibitemOpen
  \bibfield  {author} {\bibinfo {author} {\bibfnamefont {A.}~\bibnamefont
  {Barbot}}, \bibinfo {author} {\bibfnamefont {M.}~\bibnamefont {Lerbinger}},
  \bibinfo {author} {\bibfnamefont {A.}~\bibnamefont {Hernandez-Garcia}},
  \bibinfo {author} {\bibfnamefont {R.}~\bibnamefont
  {Garc{\'\i}a-Garc{\'\i}a}}, \bibinfo {author} {\bibfnamefont {M.~L.}\
  \bibnamefont {Falk}}, \bibinfo {author} {\bibfnamefont {D.}~\bibnamefont
  {Vandembroucq}}, \ and\ \bibinfo {author} {\bibfnamefont {S.}~\bibnamefont
  {Patinet}},\ }\href
  {https://journals.aps.org/pre/abstract/10.1103/PhysRevE.97.033001} {\bibfield
   {journal} {\bibinfo  {journal} {Physical Review E}\ }\textbf {\bibinfo
  {volume} {97}},\ \bibinfo {pages} {033001} (\bibinfo {year}
  {2018})}\BibitemShut {NoStop}%
\bibitem [{\citenamefont {Coslovich}(2011)}]{coslovich2011locally}%
  \BibitemOpen
  \bibfield  {author} {\bibinfo {author} {\bibfnamefont {D.}~\bibnamefont
  {Coslovich}},\ }\href
  {https://journals.aps.org/pre/abstract/10.1103/PhysRevE.83.051505} {\bibfield
   {journal} {\bibinfo  {journal} {Physical Review E}\ }\textbf {\bibinfo
  {volume} {83}},\ \bibinfo {pages} {051505} (\bibinfo {year}
  {2011})}\BibitemShut {NoStop}%
\bibitem [{\citenamefont {Malins}\ \emph
  {et~al.}(2013{\natexlab{a}})\citenamefont {Malins}, \citenamefont {Eggers},
  \citenamefont {Royall}, \citenamefont {Williams},\ and\ \citenamefont
  {Tanaka}}]{malins2013identification1}%
  \BibitemOpen
  \bibfield  {author} {\bibinfo {author} {\bibfnamefont {A.}~\bibnamefont
  {Malins}}, \bibinfo {author} {\bibfnamefont {J.}~\bibnamefont {Eggers}},
  \bibinfo {author} {\bibfnamefont {C.~P.}\ \bibnamefont {Royall}}, \bibinfo
  {author} {\bibfnamefont {S.~R.}\ \bibnamefont {Williams}}, \ and\ \bibinfo
  {author} {\bibfnamefont {H.}~\bibnamefont {Tanaka}},\ }\href
  {https://aip.scitation.org/doi/full/10.1063/1.4790515?casa_token=OGJiwU9R0ywAAAAA%3AyJd7GZwNfaPeCpNPNB25f96Ud4FPBpVedm2pBBKtFw_iaH4HDgVmF8nBV-CHIN_bhZnO9AutZBQ}
  {\bibfield  {journal} {\bibinfo  {journal} {The Journal of Chemical Physics}\
  }\textbf {\bibinfo {volume} {138}},\ \bibinfo {pages} {12A535} (\bibinfo
  {year} {2013}{\natexlab{a}})}\BibitemShut {NoStop}%
\bibitem [{\citenamefont {Malins}\ \emph
  {et~al.}(2013{\natexlab{b}})\citenamefont {Malins}, \citenamefont {Williams},
  \citenamefont {Eggers},\ and\ \citenamefont
  {Royall}}]{malins2013identification}%
  \BibitemOpen
  \bibfield  {author} {\bibinfo {author} {\bibfnamefont {A.}~\bibnamefont
  {Malins}}, \bibinfo {author} {\bibfnamefont {S.~R.}\ \bibnamefont
  {Williams}}, \bibinfo {author} {\bibfnamefont {J.}~\bibnamefont {Eggers}}, \
  and\ \bibinfo {author} {\bibfnamefont {C.~P.}\ \bibnamefont {Royall}},\
  }\href
  {https://aip.scitation.org/doi/full/10.1063/1.4832897?casa_token=r-f9UcDAPQQAAAAA%3AJzIXSwiwT8447JTEL3dwECcoiNIwCI42S_Fl8E7c5vtn_fAcTFruLexNCcYEEasHWxjjxtwO9oA}
  {\bibfield  {journal} {\bibinfo  {journal} {The Journal of Chemical Physics}\
  }\textbf {\bibinfo {volume} {139}},\ \bibinfo {pages} {234506} (\bibinfo
  {year} {2013}{\natexlab{b}})}\BibitemShut {NoStop}%
\bibitem [{\citenamefont {Tong}\ and\ \citenamefont
  {Tanaka}(2018)}]{tong2018revealing}%
  \BibitemOpen
  \bibfield  {author} {\bibinfo {author} {\bibfnamefont {H.}~\bibnamefont
  {Tong}}\ and\ \bibinfo {author} {\bibfnamefont {H.}~\bibnamefont {Tanaka}},\
  }\href {https://journals.aps.org/prx/abstract/10.1103/PhysRevX.8.011041}
  {\bibfield  {journal} {\bibinfo  {journal} {Physical Review X}\ }\textbf
  {\bibinfo {volume} {8}},\ \bibinfo {pages} {011041} (\bibinfo {year}
  {2018})}\BibitemShut {NoStop}%
\bibitem [{\citenamefont {Paret}\ \emph {et~al.}(2020)\citenamefont {Paret},
  \citenamefont {Jack},\ and\ \citenamefont {Coslovich}}]{paret2020assessing}%
  \BibitemOpen
  \bibfield  {author} {\bibinfo {author} {\bibfnamefont {J.}~\bibnamefont
  {Paret}}, \bibinfo {author} {\bibfnamefont {R.~L.}\ \bibnamefont {Jack}}, \
  and\ \bibinfo {author} {\bibfnamefont {D.}~\bibnamefont {Coslovich}},\ }\href
  {https://aip.scitation.org/doi/full/10.1063/5.0004732?casa_token=EAei_1krSYoAAAAA%3A6FPlcqEpCv9a-mF0ghj2PzSMKBQhDWNwybLQX-gmC9zJmtCr9vvIqI7JJoSavOpYT7seE-3NfMk}
  {\bibfield  {journal} {\bibinfo  {journal} {The Journal of Chemical Physics}\
  }\textbf {\bibinfo {volume} {152}},\ \bibinfo {pages} {144502} (\bibinfo
  {year} {2020})}\BibitemShut {NoStop}%
\bibitem [{\citenamefont {Boattini}\ \emph {et~al.}(2020)\citenamefont
  {Boattini}, \citenamefont {Mar{\'\i}n-Aguilar}, \citenamefont {Mitra},
  \citenamefont {Foffi}, \citenamefont {Smallenburg},\ and\ \citenamefont
  {Filion}}]{boattini2020autonomously}%
  \BibitemOpen
  \bibfield  {author} {\bibinfo {author} {\bibfnamefont {E.}~\bibnamefont
  {Boattini}}, \bibinfo {author} {\bibfnamefont {S.}~\bibnamefont
  {Mar{\'\i}n-Aguilar}}, \bibinfo {author} {\bibfnamefont {S.}~\bibnamefont
  {Mitra}}, \bibinfo {author} {\bibfnamefont {G.}~\bibnamefont {Foffi}},
  \bibinfo {author} {\bibfnamefont {F.}~\bibnamefont {Smallenburg}}, \ and\
  \bibinfo {author} {\bibfnamefont {L.}~\bibnamefont {Filion}},\ }\href
  {https://www.nature.com/articles/s41467-020-19286-8} {\bibfield  {journal}
  {\bibinfo  {journal} {Nature Communications}\ }\textbf {\bibinfo {volume}
  {11}},\ \bibinfo {pages} {1} (\bibinfo {year} {2020})}\BibitemShut {NoStop}%
\bibitem [{\citenamefont {Franz}(2005)}]{franz2005first}%
  \BibitemOpen
  \bibfield  {author} {\bibinfo {author} {\bibfnamefont {S.}~\bibnamefont
  {Franz}},\ }\href
  {https://iopscience.iop.org/article/10.1088/1742-5468/2005/04/P04001/meta}
  {\bibfield  {journal} {\bibinfo  {journal} {Journal of Statistical Mechanics:
  Theory and Experiment}\ }\textbf {\bibinfo {volume} {2005}},\ \bibinfo
  {pages} {P04001} (\bibinfo {year} {2005})}\BibitemShut {NoStop}%
\bibitem [{\citenamefont {Dzero}\ \emph {et~al.}(2005)\citenamefont {Dzero},
  \citenamefont {Schmalian},\ and\ \citenamefont
  {Wolynes}}]{dzero2005activated}%
  \BibitemOpen
  \bibfield  {author} {\bibinfo {author} {\bibfnamefont {M.}~\bibnamefont
  {Dzero}}, \bibinfo {author} {\bibfnamefont {J.}~\bibnamefont {Schmalian}}, \
  and\ \bibinfo {author} {\bibfnamefont {P.~G.}\ \bibnamefont {Wolynes}},\
  }\href {https://journals.aps.org/prb/abstract/10.1103/PhysRevB.72.100201}
  {\bibfield  {journal} {\bibinfo  {journal} {Physical Review B}\ }\textbf
  {\bibinfo {volume} {72}},\ \bibinfo {pages} {100201} (\bibinfo {year}
  {2005})}\BibitemShut {NoStop}%
\bibitem [{\citenamefont {Hocky}\ \emph {et~al.}(2014)\citenamefont {Hocky},
  \citenamefont {Coslovich}, \citenamefont {Ikeda},\ and\ \citenamefont
  {Reichman}}]{hocky2014correlation}%
  \BibitemOpen
  \bibfield  {author} {\bibinfo {author} {\bibfnamefont {G.~M.}\ \bibnamefont
  {Hocky}}, \bibinfo {author} {\bibfnamefont {D.}~\bibnamefont {Coslovich}},
  \bibinfo {author} {\bibfnamefont {A.}~\bibnamefont {Ikeda}}, \ and\ \bibinfo
  {author} {\bibfnamefont {D.~R.}\ \bibnamefont {Reichman}},\ }\href
  {https://journals.aps.org/prl/abstract/10.1103/PhysRevLett.113.157801}
  {\bibfield  {journal} {\bibinfo  {journal} {Physical Review Letters}\
  }\textbf {\bibinfo {volume} {113}},\ \bibinfo {pages} {157801} (\bibinfo
  {year} {2014})}\BibitemShut {NoStop}%
\bibitem [{\citenamefont {Charbonneau}\ \emph {et~al.}(2016)\citenamefont
  {Charbonneau}, \citenamefont {Dyer}, \citenamefont {Lee},\ and\ \citenamefont
  {Yaida}}]{charbonneau2016linking}%
  \BibitemOpen
  \bibfield  {author} {\bibinfo {author} {\bibfnamefont {P.}~\bibnamefont
  {Charbonneau}}, \bibinfo {author} {\bibfnamefont {E.}~\bibnamefont {Dyer}},
  \bibinfo {author} {\bibfnamefont {J.}~\bibnamefont {Lee}}, \ and\ \bibinfo
  {author} {\bibfnamefont {S.}~\bibnamefont {Yaida}},\ }\href
  {https://iopscience.iop.org/article/10.1088/1742-5468/2016/07/074004/meta?casa_token=Asfkn1fxQQ8AAAAA:RAG7i3PAwahzK1Lykw528F2saoMp0Yv74kxkPKKGlIgNzKVVa3sexov81ondf_w9gelPM3oIsds}
  {\bibfield  {journal} {\bibinfo  {journal} {Journal of Statistical Mechanics:
  Theory and Experiment}\ }\textbf {\bibinfo {volume} {2016}},\ \bibinfo
  {pages} {074004} (\bibinfo {year} {2016})}\BibitemShut {NoStop}%
\bibitem [{\citenamefont {Hukushima}\ and\ \citenamefont
  {Nemoto}(1996)}]{hukushima1996exchange}%
  \BibitemOpen
  \bibfield  {author} {\bibinfo {author} {\bibfnamefont {K.}~\bibnamefont
  {Hukushima}}\ and\ \bibinfo {author} {\bibfnamefont {K.}~\bibnamefont
  {Nemoto}},\ }\href {https://journals.jps.jp/doi/abs/10.1143/JPSJ.65.1604}
  {\bibfield  {journal} {\bibinfo  {journal} {Journal of the Physical Society
  of Japan}\ }\textbf {\bibinfo {volume} {65}},\ \bibinfo {pages} {1604}
  (\bibinfo {year} {1996})}\BibitemShut {NoStop}%
\bibitem [{\citenamefont {Brumer}\ and\ \citenamefont
  {Reichman}(2004)}]{brumer2004numerical}%
  \BibitemOpen
  \bibfield  {author} {\bibinfo {author} {\bibfnamefont {Y.}~\bibnamefont
  {Brumer}}\ and\ \bibinfo {author} {\bibfnamefont {D.~R.}\ \bibnamefont
  {Reichman}},\ }\href
  {https://pubs.acs.org/doi/abs/10.1021/jp037617y?casa_token=NjQyMKuHLBYAAAAA:K503VtiFXk3z_LmvO82BcDrwHwo_9lxLIW-UQMJp5tkcd1awjioPJ4YM9nSix6wPfRZNGlkXK0m8zBo}
  {\bibfield  {journal} {\bibinfo  {journal} {The Journal of Physical Chemistry
  B}\ }\textbf {\bibinfo {volume} {108}},\ \bibinfo {pages} {6832} (\bibinfo
  {year} {2004})}\BibitemShut {NoStop}%
\bibitem [{\citenamefont {Ninarello}\ \emph {et~al.}(2017)\citenamefont
  {Ninarello}, \citenamefont {Berthier},\ and\ \citenamefont
  {Coslovich}}]{ninarello2017models}%
  \BibitemOpen
  \bibfield  {author} {\bibinfo {author} {\bibfnamefont {A.}~\bibnamefont
  {Ninarello}}, \bibinfo {author} {\bibfnamefont {L.}~\bibnamefont {Berthier}},
  \ and\ \bibinfo {author} {\bibfnamefont {D.}~\bibnamefont {Coslovich}},\
  }\href {https://journals.aps.org/prx/abstract/10.1103/PhysRevX.7.021039}
  {\bibfield  {journal} {\bibinfo  {journal} {Physical Review X}\ }\textbf
  {\bibinfo {volume} {7}},\ \bibinfo {pages} {021039} (\bibinfo {year}
  {2017})}\BibitemShut {NoStop}%
\bibitem [{\citenamefont {Guiselin}\ \emph {et~al.}(2021)\citenamefont
  {Guiselin}, \citenamefont {Scalliet},\ and\ \citenamefont
  {Berthier}}]{guiselin2021microscopic}%
  \BibitemOpen
  \bibfield  {author} {\bibinfo {author} {\bibfnamefont {B.}~\bibnamefont
  {Guiselin}}, \bibinfo {author} {\bibfnamefont {C.}~\bibnamefont {Scalliet}},
  \ and\ \bibinfo {author} {\bibfnamefont {L.}~\bibnamefont {Berthier}},\
  }\href {https://arxiv.org/abs/2103.01569} {\bibfield  {journal} {\bibinfo
  {journal} {Nature Physics (in press), arXiv preprint arXiv:2103.01569}\ }
  (\bibinfo {year} {2021})}\BibitemShut {NoStop}%
\bibitem [{\citenamefont {Nos{\'e}}(1984)}]{nose1984unified}%
  \BibitemOpen
  \bibfield  {author} {\bibinfo {author} {\bibfnamefont {S.}~\bibnamefont
  {Nos{\'e}}},\ }\href
  {https://aip.scitation.org/doi/abs/10.1063/1.447334?casa_token=khUM8sZ2Y8IAAAAA:TQrdk847xvzi9AM5Bh4hGkFmn1kTV0qjmb5fFHDI1XRJvpzwm5h-fsDWMtwcqie9FziLNJC2K-c}
  {\bibfield  {journal} {\bibinfo  {journal} {The Journal of Chemical Physics}\
  }\textbf {\bibinfo {volume} {81}},\ \bibinfo {pages} {511} (\bibinfo {year}
  {1984})}\BibitemShut {NoStop}%
\bibitem [{\citenamefont {Hoover}(1985)}]{hoover1985canonical}%
  \BibitemOpen
  \bibfield  {author} {\bibinfo {author} {\bibfnamefont {W.~G.}\ \bibnamefont
  {Hoover}},\ }\href
  {https://journals.aps.org/pra/abstract/10.1103/PhysRevA.31.1695} {\bibfield
  {journal} {\bibinfo  {journal} {Physical Review A}\ }\textbf {\bibinfo
  {volume} {31}},\ \bibinfo {pages} {1695} (\bibinfo {year}
  {1985})}\BibitemShut {NoStop}%
\bibitem [{\citenamefont {Martyna}\ \emph {et~al.}(1992)\citenamefont
  {Martyna}, \citenamefont {Klein},\ and\ \citenamefont
  {Tuckerman}}]{martyna1992nose}%
  \BibitemOpen
  \bibfield  {author} {\bibinfo {author} {\bibfnamefont {G.~J.}\ \bibnamefont
  {Martyna}}, \bibinfo {author} {\bibfnamefont {M.~L.}\ \bibnamefont {Klein}},
  \ and\ \bibinfo {author} {\bibfnamefont {M.}~\bibnamefont {Tuckerman}},\
  }\href
  {https://aip.scitation.org/doi/abs/10.1063/1.463940?casa_token=t9AOBR9nuOoAAAAA:fxLhroE4byrGZOgEg-DFCZI9X1by57B19kyg8OVaPLR6i4BYOhE2iN8T_k4srW2LdT6h2JuT900}
  {\bibfield  {journal} {\bibinfo  {journal} {The Journal of Chemical Physics}\
  }\textbf {\bibinfo {volume} {97}},\ \bibinfo {pages} {2635} (\bibinfo {year}
  {1992})}\BibitemShut {NoStop}%
\bibitem [{\citenamefont {Berthier}\ \emph
  {et~al.}(2019{\natexlab{c}})\citenamefont {Berthier}, \citenamefont
  {Flenner}, \citenamefont {Fullerton}, \citenamefont {Scalliet},\ and\
  \citenamefont {Singh}}]{berthier2019efficient}%
  \BibitemOpen
  \bibfield  {author} {\bibinfo {author} {\bibfnamefont {L.}~\bibnamefont
  {Berthier}}, \bibinfo {author} {\bibfnamefont {E.}~\bibnamefont {Flenner}},
  \bibinfo {author} {\bibfnamefont {C.~J.}\ \bibnamefont {Fullerton}}, \bibinfo
  {author} {\bibfnamefont {C.}~\bibnamefont {Scalliet}}, \ and\ \bibinfo
  {author} {\bibfnamefont {M.}~\bibnamefont {Singh}},\ }\href
  {https://iopscience.iop.org/article/10.1088/1742-5468/ab1910/meta} {\bibfield
   {journal} {\bibinfo  {journal} {Journal of Statistical Mechanics: Theory and
  Experiment}\ }\textbf {\bibinfo {volume} {2019}},\ \bibinfo {pages} {064004}
  (\bibinfo {year} {2019}{\natexlab{c}})}\BibitemShut {NoStop}%
\bibitem [{\citenamefont {Martyna}\ \emph {et~al.}(1996)\citenamefont
  {Martyna}, \citenamefont {Tuckerman}, \citenamefont {Tobias},\ and\
  \citenamefont {Klein}}]{martyna1996explicit}%
  \BibitemOpen
  \bibfield  {author} {\bibinfo {author} {\bibfnamefont {G.~J.}\ \bibnamefont
  {Martyna}}, \bibinfo {author} {\bibfnamefont {M.~E.}\ \bibnamefont
  {Tuckerman}}, \bibinfo {author} {\bibfnamefont {D.~J.}\ \bibnamefont
  {Tobias}}, \ and\ \bibinfo {author} {\bibfnamefont {M.~L.}\ \bibnamefont
  {Klein}},\ }\href
  {https://www.tandfonline.com/doi/abs/10.1080/00268979600100761} {\bibfield
  {journal} {\bibinfo  {journal} {Molecular Physics}\ }\textbf {\bibinfo
  {volume} {87}},\ \bibinfo {pages} {1117} (\bibinfo {year}
  {1996})}\BibitemShut {NoStop}%
\bibitem [{\citenamefont {Frenkel}\ and\ \citenamefont
  {Smit}(2001)}]{frenkel2001understanding}%
  \BibitemOpen
  \bibfield  {author} {\bibinfo {author} {\bibfnamefont {D.}~\bibnamefont
  {Frenkel}}\ and\ \bibinfo {author} {\bibfnamefont {B.}~\bibnamefont {Smit}},\
  }\href@noop {} {\emph {\bibinfo {title} {Understanding molecular simulation:
  from algorithms to applications}}}\ (\bibinfo  {publisher} {Elsevier},\
  \bibinfo {year} {2001})\BibitemShut {NoStop}%
\bibitem [{\citenamefont {Guiselin}\ \emph
  {et~al.}(2020{\natexlab{b}})\citenamefont {Guiselin}, \citenamefont
  {Tarjus},\ and\ \citenamefont {Berthier}}]{guiselin2020overlap}%
  \BibitemOpen
  \bibfield  {author} {\bibinfo {author} {\bibfnamefont {B.}~\bibnamefont
  {Guiselin}}, \bibinfo {author} {\bibfnamefont {G.}~\bibnamefont {Tarjus}}, \
  and\ \bibinfo {author} {\bibfnamefont {L.}~\bibnamefont {Berthier}},\ }\href
  {https://aip.scitation.org/doi/full/10.1063/5.0022614?casa_token=b8PYXXa-wPoAAAAA%3Aon5_0Tckz_orsPMfVNqj28WFKu8KAkpc_HRafauoFoc0fIEmYW7RSP5gb897n-p9052z6ClQqLo}
  {\bibfield  {journal} {\bibinfo  {journal} {The Journal of Chemical Physics}\
  }\textbf {\bibinfo {volume} {153}},\ \bibinfo {pages} {224502} (\bibinfo
  {year} {2020}{\natexlab{b}})}\BibitemShut {NoStop}%
\bibitem [{\citenamefont {Ferrenberg}\ and\ \citenamefont
  {Swendsen}(1989)}]{ferrenberg1989optimized}%
  \BibitemOpen
  \bibfield  {author} {\bibinfo {author} {\bibfnamefont {A.~M.}\ \bibnamefont
  {Ferrenberg}}\ and\ \bibinfo {author} {\bibfnamefont {R.~H.}\ \bibnamefont
  {Swendsen}},\ }\href {\doibase 10.1103/PhysRevLett.63.1195} {\bibfield
  {journal} {\bibinfo  {journal} {Physical Review Letters}\ }\textbf {\bibinfo
  {volume} {63}},\ \bibinfo {pages} {1195} (\bibinfo {year}
  {1989})}\BibitemShut {NoStop}%
\bibitem [{\citenamefont {Newman}\ and\ \citenamefont
  {Barkema}(1999)}]{newman1999monte}%
  \BibitemOpen
  \bibfield  {author} {\bibinfo {author} {\bibfnamefont {M.}~\bibnamefont
  {Newman}}\ and\ \bibinfo {author} {\bibfnamefont {G.}~\bibnamefont
  {Barkema}},\ }\href@noop {} {\emph {\bibinfo {title} {Monte Carlo methods in
  statistical physics}}}\ (\bibinfo  {publisher} {Oxford University Press: New
  York, USA},\ \bibinfo {year} {1999})\BibitemShut {NoStop}%
\bibitem [{\citenamefont {Kumar}\ \emph {et~al.}(1992)\citenamefont {Kumar},
  \citenamefont {Rosenberg}, \citenamefont {Bouzida}, \citenamefont
  {Swendsen},\ and\ \citenamefont {Kollman}}]{kumar1992weighted}%
  \BibitemOpen
  \bibfield  {author} {\bibinfo {author} {\bibfnamefont {S.}~\bibnamefont
  {Kumar}}, \bibinfo {author} {\bibfnamefont {J.~M.}\ \bibnamefont
  {Rosenberg}}, \bibinfo {author} {\bibfnamefont {D.}~\bibnamefont {Bouzida}},
  \bibinfo {author} {\bibfnamefont {R.~H.}\ \bibnamefont {Swendsen}}, \ and\
  \bibinfo {author} {\bibfnamefont {P.~A.}\ \bibnamefont {Kollman}},\ }\href
  {https://onlinelibrary.wiley.com/doi/abs/10.1002/jcc.540130812?casa_token=9j_GLYjxrCIAAAAA:5-1ccMmJLZgZuGFYxiHB5nN7KMqyMvSSjnJp3BxMXPsXQCa0ZkFfOdNBrP9nXgclWo5aSbUJh6GhpxGp}
  {\bibfield  {journal} {\bibinfo  {journal} {Journal of Computational
  Chemistry}\ }\textbf {\bibinfo {volume} {13}},\ \bibinfo {pages} {1011}
  (\bibinfo {year} {1992})}\BibitemShut {NoStop}%
\bibitem [{\citenamefont {Tolman}(1949)}]{tolman1949effect}%
  \BibitemOpen
  \bibfield  {author} {\bibinfo {author} {\bibfnamefont {R.~C.}\ \bibnamefont
  {Tolman}},\ }\href
  {https://aip.scitation.org/doi/abs/10.1063/1.1747247?casa_token=sqeKhAZOllUAAAAA:PBbw5nDqMzN974kP_OOrrb2kFqTeoufFDzkkSeyjsook-cD8XZS5r-VkjG9uAFVTT91J3ngK5rs}
  {\bibfield  {journal} {\bibinfo  {journal} {The Journal of Chemical Physics}\
  }\textbf {\bibinfo {volume} {17}},\ \bibinfo {pages} {333} (\bibinfo {year}
  {1949})}\BibitemShut {NoStop}%
\bibitem [{\citenamefont {Grinstein}\ and\ \citenamefont
  {Ma}(1983)}]{grinstein1983surface}%
  \BibitemOpen
  \bibfield  {author} {\bibinfo {author} {\bibfnamefont {G.}~\bibnamefont
  {Grinstein}}\ and\ \bibinfo {author} {\bibfnamefont {S.-K.}\ \bibnamefont
  {Ma}},\ }\href
  {https://journals.aps.org/prb/abstract/10.1103/PhysRevB.28.2588} {\bibfield
  {journal} {\bibinfo  {journal} {Physical Review B}\ }\textbf {\bibinfo
  {volume} {28}},\ \bibinfo {pages} {2588} (\bibinfo {year}
  {1983})}\BibitemShut {NoStop}%
\bibitem [{\citenamefont {Stevenson}\ \emph {et~al.}(2006)\citenamefont
  {Stevenson}, \citenamefont {Schmalian},\ and\ \citenamefont
  {Wolynes}}]{stevenson2006shapes}%
  \BibitemOpen
  \bibfield  {author} {\bibinfo {author} {\bibfnamefont {J.~D.}\ \bibnamefont
  {Stevenson}}, \bibinfo {author} {\bibfnamefont {J.}~\bibnamefont
  {Schmalian}}, \ and\ \bibinfo {author} {\bibfnamefont {P.~G.}\ \bibnamefont
  {Wolynes}},\ }\href {https://www.nature.com/articles/nphys261} {\bibfield
  {journal} {\bibinfo  {journal} {Nature Physics}\ }\textbf {\bibinfo {volume}
  {2}},\ \bibinfo {pages} {268} (\bibinfo {year} {2006})}\BibitemShut {NoStop}%
\bibitem [{\citenamefont {Biroli}\ and\ \citenamefont
  {Cammarota}(2017)}]{biroli2017fluctuations}%
  \BibitemOpen
  \bibfield  {author} {\bibinfo {author} {\bibfnamefont {G.}~\bibnamefont
  {Biroli}}\ and\ \bibinfo {author} {\bibfnamefont {C.}~\bibnamefont
  {Cammarota}},\ }\href
  {https://journals.aps.org/prx/abstract/10.1103/PhysRevX.7.011011} {\bibfield
  {journal} {\bibinfo  {journal} {Physical Review X}\ }\textbf {\bibinfo
  {volume} {7}},\ \bibinfo {pages} {011011} (\bibinfo {year}
  {2017})}\BibitemShut {NoStop}%
\bibitem [{\citenamefont {Wyart}\ and\ \citenamefont
  {Cates}(2017)}]{PhysRevLett.119.195501}%
  \BibitemOpen
  \bibfield  {author} {\bibinfo {author} {\bibfnamefont {M.}~\bibnamefont
  {Wyart}}\ and\ \bibinfo {author} {\bibfnamefont {M.~E.}\ \bibnamefont
  {Cates}},\ }\href {\doibase 10.1103/PhysRevLett.119.195501} {\bibfield
  {journal} {\bibinfo  {journal} {Phys. Rev. Lett.}\ }\textbf {\bibinfo
  {volume} {119}},\ \bibinfo {pages} {195501} (\bibinfo {year}
  {2017})}\BibitemShut {NoStop}%
\bibitem [{\citenamefont {Berthier}\ \emph
  {et~al.}(2019{\natexlab{d}})\citenamefont {Berthier}, \citenamefont {Biroli},
  \citenamefont {Bouchaud},\ and\ \citenamefont
  {Tarjus}}]{doi:10.1063/1.5086509}%
  \BibitemOpen
  \bibfield  {author} {\bibinfo {author} {\bibfnamefont {L.}~\bibnamefont
  {Berthier}}, \bibinfo {author} {\bibfnamefont {G.}~\bibnamefont {Biroli}},
  \bibinfo {author} {\bibfnamefont {J.-P.}\ \bibnamefont {Bouchaud}}, \ and\
  \bibinfo {author} {\bibfnamefont {G.}~\bibnamefont {Tarjus}},\ }\href
  {\doibase 10.1063/1.5086509} {\bibfield  {journal} {\bibinfo  {journal} {The
  Journal of Chemical Physics}\ }\textbf {\bibinfo {volume} {150}},\ \bibinfo
  {pages} {094501} (\bibinfo {year} {2019}{\natexlab{d}})}\BibitemShut
  {NoStop}%
\bibitem [{\citenamefont {Bramwell}\ \emph {et~al.}(1998)\citenamefont
  {Bramwell}, \citenamefont {Holdsworth},\ and\ \citenamefont
  {Pinton}}]{bramwell1998universality}%
  \BibitemOpen
  \bibfield  {author} {\bibinfo {author} {\bibfnamefont {S.}~\bibnamefont
  {Bramwell}}, \bibinfo {author} {\bibfnamefont {P.}~\bibnamefont
  {Holdsworth}}, \ and\ \bibinfo {author} {\bibfnamefont {J.-F.}\ \bibnamefont
  {Pinton}},\ }\href {https://www.nature.com/articles/25083} {\bibfield
  {journal} {\bibinfo  {journal} {Nature}\ }\textbf {\bibinfo {volume} {396}},\
  \bibinfo {pages} {552} (\bibinfo {year} {1998})}\BibitemShut {NoStop}%
\bibitem [{\citenamefont {Bramwell}\ \emph {et~al.}(2000)\citenamefont
  {Bramwell}, \citenamefont {Christensen}, \citenamefont {Fortin},
  \citenamefont {Holdsworth}, \citenamefont {Jensen}, \citenamefont {Lise},
  \citenamefont {L{\'o}pez}, \citenamefont {Nicodemi}, \citenamefont {Pinton},\
  and\ \citenamefont {Sellitto}}]{bramwell2000universal}%
  \BibitemOpen
  \bibfield  {author} {\bibinfo {author} {\bibfnamefont {S.}~\bibnamefont
  {Bramwell}}, \bibinfo {author} {\bibfnamefont {K.}~\bibnamefont
  {Christensen}}, \bibinfo {author} {\bibfnamefont {J.-Y.}\ \bibnamefont
  {Fortin}}, \bibinfo {author} {\bibfnamefont {P.}~\bibnamefont {Holdsworth}},
  \bibinfo {author} {\bibfnamefont {H.}~\bibnamefont {Jensen}}, \bibinfo
  {author} {\bibfnamefont {S.}~\bibnamefont {Lise}}, \bibinfo {author}
  {\bibfnamefont {J.}~\bibnamefont {L{\'o}pez}}, \bibinfo {author}
  {\bibfnamefont {M.}~\bibnamefont {Nicodemi}}, \bibinfo {author}
  {\bibfnamefont {J.-F.}\ \bibnamefont {Pinton}}, \ and\ \bibinfo {author}
  {\bibfnamefont {M.}~\bibnamefont {Sellitto}},\ }\href
  {https://journals.aps.org/prl/abstract/10.1103/PhysRevLett.84.3744}
  {\bibfield  {journal} {\bibinfo  {journal} {Physical Review Letters}\
  }\textbf {\bibinfo {volume} {84}},\ \bibinfo {pages} {3744} (\bibinfo {year}
  {2000})}\BibitemShut {NoStop}%
\bibitem [{\citenamefont {Bramwell}\ \emph {et~al.}(2001)\citenamefont
  {Bramwell}, \citenamefont {Fortin}, \citenamefont {Holdsworth}, \citenamefont
  {Peysson}, \citenamefont {Pinton}, \citenamefont {Portelli},\ and\
  \citenamefont {Sellitto}}]{bramwell2001magnetic}%
  \BibitemOpen
  \bibfield  {author} {\bibinfo {author} {\bibfnamefont {S.}~\bibnamefont
  {Bramwell}}, \bibinfo {author} {\bibfnamefont {J.-Y.}\ \bibnamefont
  {Fortin}}, \bibinfo {author} {\bibfnamefont {P.}~\bibnamefont {Holdsworth}},
  \bibinfo {author} {\bibfnamefont {S.}~\bibnamefont {Peysson}}, \bibinfo
  {author} {\bibfnamefont {J.-F.}\ \bibnamefont {Pinton}}, \bibinfo {author}
  {\bibfnamefont {B.}~\bibnamefont {Portelli}}, \ and\ \bibinfo {author}
  {\bibfnamefont {M.}~\bibnamefont {Sellitto}},\ }\href
  {https://journals.aps.org/pre/abstract/10.1103/PhysRevE.63.041106} {\bibfield
   {journal} {\bibinfo  {journal} {Physical Review E}\ }\textbf {\bibinfo
  {volume} {63}},\ \bibinfo {pages} {041106} (\bibinfo {year}
  {2001})}\BibitemShut {NoStop}%
\bibitem [{\citenamefont {Coslovich}\ and\ \citenamefont
  {Jack}(2016)}]{coslovich2016structure}%
  \BibitemOpen
  \bibfield  {author} {\bibinfo {author} {\bibfnamefont {D.}~\bibnamefont
  {Coslovich}}\ and\ \bibinfo {author} {\bibfnamefont {R.~L.}\ \bibnamefont
  {Jack}},\ }\href
  {https://iopscience.iop.org/article/10.1088/1742-5468/2016/07/074012/meta?casa_token=F1xbMdzA7noAAAAA:MfhPvZYyPPLAt_wadAwHy5UUP_4OIHQN4H1aMKUA1ewJCAFd1A-g1TWBLB2OBRLGpC3vT9Z56X4}
  {\bibfield  {journal} {\bibinfo  {journal} {Journal of Statistical Mechanics:
  Theory and Experiment}\ }\textbf {\bibinfo {volume} {2016}},\ \bibinfo
  {pages} {074012} (\bibinfo {year} {2016})}\BibitemShut {NoStop}%
\bibitem [{\citenamefont {Adam}\ and\ \citenamefont
  {Gibbs}(1965)}]{adam1965temperature}%
  \BibitemOpen
  \bibfield  {author} {\bibinfo {author} {\bibfnamefont {G.}~\bibnamefont
  {Adam}}\ and\ \bibinfo {author} {\bibfnamefont {J.~H.}\ \bibnamefont
  {Gibbs}},\ }\href
  {https://aip.scitation.org/doi/abs/10.1063/1.1696442?casa_token=znr_aA9gGWkAAAAA:kcKfje8nm8Vc3zaT01mXBraj70qED1KgGdb7Gos7bI7ukfF2-i4gwmBlVSFtE5qr92QZvCCqruk}
  {\bibfield  {journal} {\bibinfo  {journal} {The Journal of Chemical Physics}\
  }\textbf {\bibinfo {volume} {43}},\ \bibinfo {pages} {139} (\bibinfo {year}
  {1965})}\BibitemShut {NoStop}%
\end{thebibliography}%

\end{document}